\documentclass[
twocolumn,
amsmath,amssymb,
aps,
prl,
floatfix,
]{revtex4-2}

\usepackage{graphicx}
\usepackage{dcolumn}
\usepackage{bm}
\usepackage{xcolor}
\usepackage{physics}
\usepackage{slashed}
\usepackage[normalem]{ulem}
\usepackage{tikz-feynman}
\tikzfeynmanset{compat=1.1.0}
\usepackage[colorlinks=true,
linkcolor=blue,
citecolor=blue,
urlcolor=blue]{hyperref}

\begin{document}

\title{Defect-Mediated Conversion: Dark Matter from Cosmological Domain-Wall Scattering}

\author{João Paulo Pinheiro}
\email{joaopaulo.pinheiro@sjtu.edu.cn}
\affiliation{State Key Laboratory of Dark Matter Physics, Tsung-Dao Lee Institute \& School of Physics and Astronomy, Shanghai Jiao Tong University, Shanghai 200240, China}
\affiliation{Key Laboratory for Particle Astrophysics and Cosmology (MOE) \& Shanghai Key Laboratory for Particle Physics and Cosmology, Shanghai Jiao Tong University, Shanghai 200240, China}

\author{Mohamed Younes Sassi}
\email{mohamedyounessassi@sjtu.edu.cn}
\affiliation{State Key Laboratory of Dark Matter Physics, Tsung-Dao Lee Institute \& School of Physics and Astronomy, Shanghai Jiao Tong University, Shanghai 200240, China}
\affiliation{Key Laboratory for Particle Astrophysics and Cosmology (MOE) \& Shanghai Key Laboratory for Particle Physics and Cosmology, Shanghai Jiao Tong University, Shanghai 200240, China}

\date{\today}

\begin{abstract}
We propose a new mechanism for dark matter genesis, \emph{Defect-Mediated
Conversion} (DMC).
Cosmological domain walls can host a scalar condensate in their core acting as a mixing 
portal between the Standard Model and a dark sector: thermal-bath fermions
crossing the wall are partially and non-thermally converted into dark particles.
The conversion probability is governed by a single dimensionless mixing area
built from the wall profile, and the yield by the area density of the network.
For a light dark particle the probability is independent of both the incident
energy and the dark mass, and the relic abundance collapses onto a single
mass--coupling relation running from the keV Lyman-$\alpha$ floor to the
wall-formation scale.
For a heavy one, conversion turns the energy of a single incident fermion into
the dark mass, populating dark sectors with masses exceeding both the bath
temperature and the mediator mass.
The same portal also sources ordinary freeze-in, but DMC, boosted by the
coherent wall-area enhancement, dominates it by one to two orders of magnitude
at fixed couplings and remains operative in the heavy regime, where freeze-in
shuts off.
\end{abstract}

\maketitle


\textbf{Dark matter production via a Defect-Mediated Conversion mechanism}
-- Every dark-matter candidate must explain how it was
produced~\cite{Bertone:2004pz,Arcadi:2017kky}.
Thermal freeze-out~\cite{Lee:1977ua,Scherrer:1985zt},
freeze-in~\cite{Hall:2009bx,Bernal:2017kxu}, and
misalignment~\cite{Preskill:1982cy,Abbott:1982af,Dine:1982ah} all connect the
relic abundance to a coupling that simultaneously fixes a detectable signal or a
narrow viable mass window.
Freeze-in, for instance, builds up the dark abundance through the slow decay of
a mediator $\chi$ held in equilibrium with the bath.
Production is efficient only while that mediator is abundant, shuts off
exponentially once $T$ falls below its mass, and limits the reachable dark mass at
the mediator mass itself.

\begin{figure}[h!]
\centering
\includegraphics[width=0.9\linewidth]{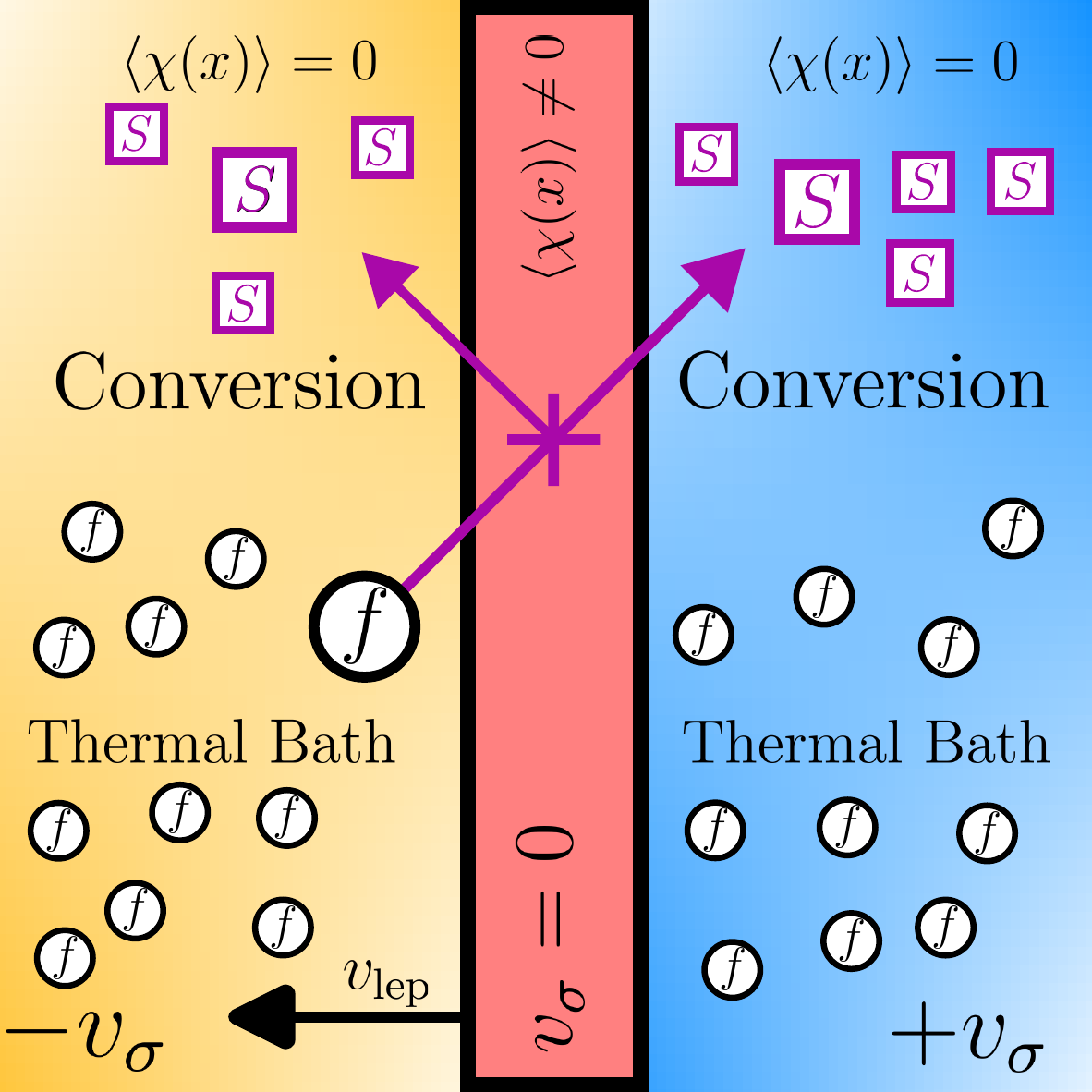}
\caption{The Defect-Mediated Conversion mechanism. Blue: a domain populated by the 
VEV $+v_\sigma$. Orange: a domain populated by the VEV
$-v_\sigma$. Red band: the domain wall transition region. Black circles: Standard Model fermions
$f$ of the thermal bath. Purple squares: dark particles $S$. Purple arrows:
transmission and reflection channels.}
\label{fig:scattering_DW}
\end{figure}
What all of these DM mechanisms share is that production happens homogeneously, everywhere in
the plasma at once.
Alternatively, if a cosmological defect sweeps the bath, the crossing itself can set the
abundance.
This has been exploited at the walls of a first-order phase transition, either
by filtering a pre-existing thermal population, so that only particles energetic
enough to cross survive~\cite{Baker:2019ndr,Chway:2019kft}, or by splitting a
bath quantum into a pair of heavy states~\cite{Azatov:2021ifm}.
All require a first-order transition, usually with substantial supercooling, and
in all of them the wall is transient, leaving a strong stochastic
gravitational-wave background as the unavoidable signature.

Domain walls, on the other hand, need none of this. They are generic relics of spontaneously broken discrete
symmetries~\cite{Zeldovich:1974uw,Saikawa:2017hiv}, form at any type of phase transitions, and persist as a scaling network over many Hubble times.
We show for the first time (to our knowledge) that such a network can be the engine of DM genesis,
rather than a means of sweeping unwanted relics out of the
plasma~\cite{Dvali:1997sa}.
A $\chi$ condensate trapped in the wall core opens a portal on the wall itself, and
bath fermions crossing it are partially converted into a dark fermion $S$.
The enhancement coming from the coherent
area of the network integrated from the formation temperature of the localized condensate $T_{\rm onset}$ down to the annihilation
temperature $T_{\rm ann}$.

Two features distinguish this from freeze-in through the same portal.
Where the mediator decay is open, $M_S<m_{\chi}$, both channels are linear in
$\sum_\alpha|y_\alpha|^{2}$, so the comparison is a pure number.
Wall crossing stays active down to the annihilation temperature $T_{\rm ann}$ of
the network.
The coherent wall-area enhancement then makes it dominate by a factor
$30$--$100$ across our benchmarks (SM Sec.~\ref{sec:freezein}).
Above $M_S=m_{\chi}$ the comparison is not a factor but a change of scaling.
The decay closes and freeze-in survives only through $\ell^+\ell^-\to S\bar S$,
quadratic in the coupling and requiring \emph{two} bath particles above
threshold.
The wall converts the energy of a single incident fermion into the dark
mass.
DMC therefore populates dark sectors heavier than both the plasma temperature
and the mediator, where freeze-in is heavily suppressed.

\paragraph{The mechanism.}
After a discrete symmetry breaks spontaneously at $T\sim T_{\rm form}$, causally
disconnected domains settle into the two degenerate vacua
$\langle\sigma\rangle=\pm v_\sigma$, separated by walls in which $\sigma$ passes
through zero (Fig.~\ref{fig:scattering_DW}).
A second scalar $\chi$ carrying no vev outside can acquire a condensate localized in
the core, opening the chiral portal $\mathcal{L}\supset y\,\chi(x)\,\bar f S$ between a
bath fermion $f=\ell_\alpha$ and $S$, generating an effective mixing between $f$ and $S$ only on the defect. 
As the wall (velocity $v_{\rm dw}$) sweeps the plasma, a fraction of the bath is
converted into a non-thermal population of dark sector particles. 
Because this is the only coupling of $S$ to the SM it never thermalizes, and
overproduction is avoided precisely because the portal is tiny.
Two things control the yield: the strength of the conversion, carried by a
single dimensionless combination of the wall profile, and the flux of bath
leptons with $E_\ell>M_S$ reaching the wall.
It is the energy of a single incident lepton, neither the plasma temperature nor
the mediator mass $m_\chi$, that bounds the dark mass which can be produced.

\textbf{Toy model} --
We extend the SM by three fields, all $\mathrm{SU(2)}_L$ singlets: a real scalar
$\sigma$ ($Y=0$), odd under a discrete $\mathbb{Z}_2$ and neutral under
$U(1)_D$; a charged scalar $h^+$ ($Y=+1$) playing the role of $\chi$, $\mathbb{Z}_2$-even and carrying dark
charge $U(1)_D=-1$; and a Dirac fermion $S$ ($Y=0$),
$\mathbb{Z}_2$-even with $U(1)_D=+1$ and lepton number $L=+1$.
The SM Higgs doublet $\Phi$ is $\mathbb{Z}_2$-even and dark-neutral.
The $\mathbb{Z}_2$ under which $\sigma\to-\sigma$ is broken spontaneously by
$\langle\sigma\rangle=\pm v_\sigma$, so the two vacua are degenerate and
separated by domain walls, while the exact $U(1)_D$ keeps $S$ stable. 

\paragraph*{Dark sector and portal.}
The only renormalizable coupling of $S$ is
\begin{equation}
\mathcal{L}\supset
- y_\alpha\,h^+\,\overline{\ell_{R\alpha}^c}\,S
- M_S\,\bar S S
+\mathrm{h.c.},
\label{eq:Ldark}
\end{equation}
with $M_S$ a \emph{free} vector-like Dirac mass.
The charge assignments forbid $h^+\bar L L$ (the Zee vertex~\cite{Zee:1980ai})
and any $S$--$\nu$ coupling, so the portal generates \emph{no} neutrino mass and
no $0\nu\beta\beta$.

\paragraph*{Charged condensate in the core.}
The scalar potential is
\begin{align}
\notag V(\Phi, \sigma, h^+) &= V_{\rm SM}(\Phi) + \tfrac{1}{2}\mu^2_+ |h^+|^2
+\tfrac{1}{2}\mu^2_\sigma \sigma^2+ \tfrac{1}{4}\lambda_\sigma \sigma^4\\ \notag
& + \tfrac{1}{4}\lambda_{+} |h^+|^4 +\lambda_{\Phi+}\,|h^+|^2\,|\Phi|^2 + \lambda_{\sigma+}\,\sigma^2\,|h^+|^2 \\
& + \lambda_{\Phi\sigma}\,\sigma^2\,|\Phi|^2 + \epsilon\,\sigma^3.
\end{align}
To simplify the analysis we neglect the coupling between the SM Higgs $\Phi$ and
the new scalars, setting $\lambda_{\Phi\sigma},\lambda_{\Phi+}\approx0$.
The small $\mathbb{Z}_2$-breaking term $\epsilon\sigma^3$ lifts the degeneracy
between $\pm v_\sigma$ and drives the network to annihilate at $T_{\rm ann}$,
well before it could come to dominate the energy density of the universe
(SM Sec.~\ref{sec:walldom})~\cite{Sikivie:1982qv,Saikawa:2017hiv}.
Requiring the network to form bounds the induced vacuum energy bias to \cite{Saikawa:2017hiv}
$V_{\rm bias}<0.795\,V_0$, which translates into un upper bound
$\epsilon<\big(0.795/(2-0.795)\big)\,(4\mu^2_++\lambda_\sigma v^2_\sigma)/(4v_\sigma)$.
In the remainder of this work, and in order to simplify the analysis, we neglect the effect of $\epsilon\sigma^3$ on the
scalar field profiles and retain only its role in annihilating the network at $T_{\rm ann}$.

The possibility of generating a condensate of $h^+$ inside the wall depends on the sign of its effective mass,
\begin{equation}
M_+^2(x)=\tfrac{1}{2}\mu^2_++\lambda_{\sigma+}\sigma^2(x).
\end{equation}
Far from the wall, with $\mu^2_+<0$ and $\lambda_{\sigma+}>0$, $M^2_+(x)$ is positive, keeping $h^+$ from
condensing in vacuum, but it can turn tachyonic at the
core, $M_+^2(0)=\tfrac{1}{2}\mu^2_+<0$, where $\sigma(0) = 0$.
A non-zero condensate also costs gradient energy, so a tachyonic core is
necessary but not sufficient: the instability is reached only when
$\mu_+^2<-\tfrac23\lambda_{\sigma+}v_\sigma^2$, derived in SM
Sec.~\ref{sec:thermal}.
Solving the coupled scalar equations of motion (SM, Sec.~\ref{sec:dw}) then
yields a localized condensate
\begin{equation}
\langle h^+(x)\rangle=v_+(x)\neq0\ \ (\text{core}),\qquad
\langle h^+\rangle_{\rm bulk}=0,
\label{eq:condensate}
\end{equation}
which breaks $\mathrm{U(1)}_{D}$ and, after the electroweak phase transition,
also $U(1)_{\rm em}$ \emph{only inside the core}.
In SM Sec.~\ref{sec:dw} we derive the window of $m_{h^+}$ admitting a stable,
bounded condensate for $\lambda_{\Phi\sigma},\lambda_{\Phi+}\approx0$, and in SM
Sec.~\ref{sec:thermal} we discuss the effect of the LO thermal corrections to
$\mu^2_+(T)$ and $\mu^2_{\sigma}(T)$ on the wall and condensate profiles.

Since these thermal masses turn negative at different temperatures, the
condensate portal does not form together with the walls, and this delay orders
the cosmological history of the mechanism.
Below $T_c$ the field $\sigma$ acquires a VEV and the network is laid down, but
$\mu^2_+(T)$ is still positive for the benchmarks of Table~\ref{tab:bench}, so
no condensate forms.
It turns negative at $T^+_c<T_c$, and even then the solution $v_+(x)=0$ remains
stable.
The instability that triggers $v_+(0)\neq0$ is reached only at $T_{\rm onset}$,
where $\mu_+^2(T_{\rm onset})<-\tfrac23\lambda_{\sigma+}v^2_\sigma(T_{\rm onset})$.
Conversion is therefore active only in the window
$T_{\rm ann}<T<T_{\rm onset}$, whereas the initial comoving wall area is set
earlier, at $T_{\rm form}=T_c$.
We take the phase transition to be second order, so there is no separate
supercooled formation temperature and $T_{\rm form}$ coincides with $T_c$.
We evaluate the condensate at $T_{\rm eval}=0.9\,T_{\rm onset}$, where $v_+(0)$
has reached a sizable fraction of its $T=0$ value but is still below it.
Since $v_+$ grows monotonically as the universe cools, this is a conservative
choice (Table~\ref{tab:bench}).

The consequence of a non-zero condensate inside the wall is that the Yukawa of
Eq.~\eqref{eq:Ldark} realizes the mixing portal
$y_\alpha v_+(x)\,\overline{\ell_{R\alpha}^c}S$, with
$\langle\chi\rangle_{\rm wall}=v_+$, which vanishes far from the wall.

\begin{table}[t]
\caption{Benchmark points BP1--BP6, all satisfying the thin-wall requirement.
The mixing area $\mathcal{S}$ is evaluated at $T_{\rm eval}=0.9\,T_{\rm onset}$
and at $T=0$.
Quartic couplings, condensate values and $T_c$ are given in SM
Table~\ref{tab:bench_full}.}
\begin{ruledtabular}
\begin{tabular}{lcccc}
& $v_\sigma$ [GeV] & $m_{h^+}$ [GeV] & $T_{\rm onset}$ [GeV] & $\mathcal{S}(T_{\rm eval})$ \\
\hline
BP1 & 300   & 523  & 263   & 0.54 \\
BP2 & 500   & 542  & 463   & 0.42 \\
BP3 & 800   & 579  & 762   & 0.33 \\
BP4 & 2000  & 789  & 2545  & 0.20 \\
BP5 & 5000  & 2399 & 5691  & 0.47 \\
BP6 & 10000 & 4747 & 11449 & 0.47 \\
\end{tabular}
\end{ruledtabular}
\label{tab:bench}
\end{table}

\textbf{Conversion probability and its universality} --
When the wall core supports a localized condensate $v_+(x)$ vanishing as
$|x|\to\infty$, the portal induces a space-dependent Dirac mass
$M(x)=y_\alpha v_+(x)$ connecting $\ell_{R\alpha}$ to $S$.
Solving the one-dimensional Dirac equation in this background using a thin-wall approximation (SM,
Sec.~\ref{sec:dirac} and \cite{Sassi:2023cqp}), a thermal bath lepton of energy
$E_\ell$ crossing the wall converts with probability
\begin{equation}
P_{\ell_\alpha\to S}(E_\ell,k_\alpha,M_S)=
\frac{8\,g_d\,(g_d^{2}+1)\sinh^{2}k_\alpha}
{\big[(g_d-1)^{2}-(g_d+1)^{2}\cosh k_\alpha\big]^{2}},
\label{eq:Pfull}
\end{equation}
where $g_d=\sqrt{E_\ell^{2}-M_S^{2}}/(E_\ell+M_S)$ and
$k_\alpha(T)=y_\alpha\,\mathcal{S}(T)$ with
\begin{equation}
\mathcal{S}(T)\equiv\!\int\!dx\; v_+(x,T)
\simeq v_+(0,T)\,\delta_{\rm wall}(T)
=\frac{2\,v_+(0,T)}{m_\sigma(T)}
\label{eq:Sdef}
\end{equation}
the dimensionless mixing area accumulated across the wall, with
$\delta_{\rm wall}\simeq2/m_\sigma(T)$ the wall width.
In what follows we neglect the SM lepton masses, since $m_\ell\ll T$ throughout
the temperature range of interest.

\begin{figure}[h]
\centering
\includegraphics[width=0.7\linewidth]{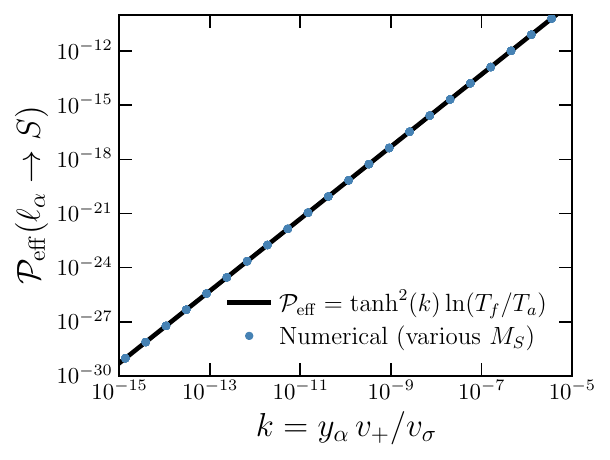}
\caption{Conversion probability $\mathcal{P}(\ell_\alpha\to S)$ as a function of
the mixing area $k=y_\alpha\mathcal{S}$, Eq.~\eqref{eq:Sdef}.
Black line: $\tanh^2(k)$.
Blue points: numerical solution of the coupled Dirac system,
Eq.~\eqref{eq:systemdirac}, for randomly drawn $M_S\ll E_\ell$.}
\label{fig:prob_conv_sm}
\end{figure}

For a light dark sector, $M_S\ll E_\ell$, the threshold variable $g_d\to1$ and
Eq.~\eqref{eq:Pfull} collapses to
\begin{equation}
P_{\ell_\alpha\to S}\;\longrightarrow\;\tanh^{2}\!\big(k_\alpha(T)\big),
\label{eq:Pgeneric}
\end{equation}
independent of the incident energy \emph{and} of $M_S$: the wall coherently
rotates $(\ell_\alpha,S)$ by the mixing $\int\!M(x)\,dx=k_\alpha$.
With no competing scale left, the exact numerical solution of the coupled Dirac
system reproduces $\tanh^{2}(k)$ with no residual dependence on $E_\ell$ or
$M_S$ (SM, Sec.~\ref{sec:dirac}, Fig.~\ref{fig:prob_conv_sm}).
The entire relativistic bath then converts at a single probability, and the flux
through the wall is simply $n_\ell\propto T^{3}$.

For a heavy dark sector, $M_S\gtrsim T$, only leptons with $E_\ell>M_S$ convert,
and Eq.~\eqref{eq:Pfull} must be integrated against the Fermi-Dirac distribution
of the bath.
The tail above threshold is never empty, so the mechanism stays operative for
$M_S$ well above $T_{\rm onset}$ and above the mediator mass $m_{h^+}$, a regime
we compare with freeze-in below.

\begin{figure}[t]
\centering
\includegraphics[width=1\linewidth]{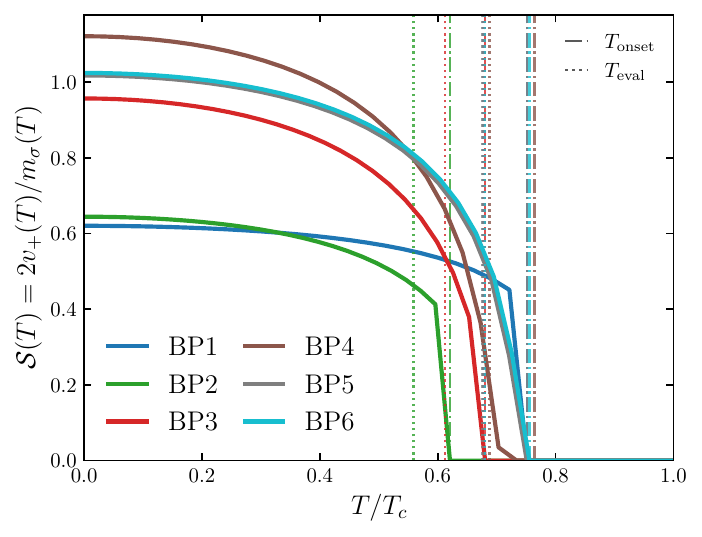}
\caption{Mixing area $\mathcal{S}(T)$, Eq.~\eqref{eq:Sdef}, for benchmarks
BP1--BP6 (Table~\ref{tab:bench}) as a function of $T/T_c$, drawn in blue, green,
red, brown, gray and cyan, respectively.
Dotted vertical lines: $T_{\rm onset}$.
Dot-dashed vertical lines: $T_{\rm eval}$.}
\label{fig:S_T}
\end{figure}

All dependence on the dimensionful parameters enters through $\mathcal{S}$,
whose value during the conversion window is shown in Fig.~\ref{fig:S_T}.
Above $T_{\rm onset}$ the portal is closed, $\mathcal{S}=0$.
It switches on at $T_{\rm onset}$ and saturates near
$\mathcal{S}(0)=2v_+(0)/m_\sigma^{0}$ at low temperatures.
For thin walls both quantities in Eq.~\eqref{eq:Sdef} are linear in the
symmetry-breaking scale, $v_+(0)\propto v_\sigma$ and
$m_\sigma\propto\sqrt{\lambda_\sigma}\,v_\sigma$, so $v_\sigma$ cancels in
their ratio and $\mathcal{S}$ is a function of the quartics
$(\lambda_\sigma,\lambda_{\sigma+},\lambda_+)$ alone.
The portal is therefore bounded from above by tree-level perturbativity,
$\lambda_i\lesssim4\pi$~\cite{Lee:1977eg}, rather than by any mass scale
(SM Sec.~\ref{sec:portalnum}).
We do not freeze the portal at a single temperature: the coupled scalar
equations of motion are solved at each $T$ across the conversion window,
giving $\mathcal{S}(T)$, which is then integrated numerically
(SM Sec.~\ref{sec:portalnum}).
Finally, the Yukawas selected by $\Omega_Sh^{2}=0.12$ give $k_\alpha\ll1$
throughout, so $\tanh^{2}k\to k^{2}$ and the portal factorizes out of the
thermal history.

\textbf{Relic abundance: dark matter at many scales} --
Apart from the conversion probability, only two ingredients enter the generated abundance: the comoving area swept by the walls,
and the flux of bath leptons through it, integrated from $T_{\rm onset}$ down to
$T_{\rm ann}$.

The area density of the network redshifts as~\cite{Vachaspati:2006zz}
\begin{equation}
\mathcal{A}(T)\approx\mathcal{A}_0\left(\frac{t_{\rm form}}{t}\right)
=\mathcal{A}_0\left(\frac{T}{T_c}\right)^{2},
\label{eq:area}
\end{equation}
with $\mathcal{A}_0=N_{\rm dw}\xi_0^{2}/H^{-3}\approx1/\xi_0$ the area density
at formation, $N_{\rm dw}\approx H^{-3}/\xi_0^{3}$ the number of walls per
Hubble volume and $\xi_0$ the correlation length of $\sigma$ at formation, set by causality at
the transition~\cite{Kibble:1976sj}.
Fixing $\mathcal{A}(T)$ precisely requires simulating
the network from formation to annihilation. 
Several effects are then relevant: thermal flipping of small domains after formation; 
friction from the sizable reflection of $h^\pm$ off the wall (and of hyperfield $B$, later the photon, once
$\langle h^+\rangle\neq0$ in the
core)~\cite{Vilenkin:1981zs,Battye:2021dyq,Blasi:2022ayo}, which delays the onset of scaling and makes  $\mathcal{A}(T)$ drop more slowly, enhancing the conversion.

The area density at formation is however bounded from above, since walls cannot be packed more densely than
their own thickness, $\xi_0\gtrsim\delta_{\rm wall}\sim1/m_\sigma^{0}$, and we
parametrize it as
\begin{equation}
\mathcal{A}_0=\gamma_{w}\,m_\sigma^{0},\qquad\gamma_w\in(0,1],
\label{eq:gamma}
\end{equation}
with $m_\sigma^{0}$ the tree-level, temperature-independent mass and $\gamma_{w}$ an
area-efficiency factor absorbing the departure of a realistic network from the
maximally packed configuration of one wall per correlation length at formation.
Causality alone gives $\gamma_{w}\le1$, and a network that coarsens beyond its own
thickness has $\gamma_{w}<1$.
The two roles of $m_\sigma$ must be kept apart: the thermal $m_\sigma(T)$ in
Eq.~\eqref{eq:Sdef} sets the wall width and hence the coherent mixing region
while the portal is open, whereas the tree-level $m_\sigma^{0}$ in
Eq.~\eqref{eq:gamma} sets the minimal correlation length at formation and hence
the maximal initial area density.

Since conversion is a local scattering off a moving wall, the source term is the
wall area times the momentum-resolved lepton flux through it,
\begin{equation}
\dot n_S+3Hn_S=\mathcal{A}(T)\,v_{\rm lep}\sum_\alpha
\frac{g_{\rm dof}}{2\pi^{2}}\int_{M_S}^{\infty}\!\!dp\;
\frac{p^{2}}{e^{p/T}+1}\;P_{\ell_\alpha\to S},
\label{eq:source_master}
\end{equation}
with $g_{\rm dof}=2$ per flavor.
The outgoing
$S$ inherits the incident energy and can only be produced for $E_\ell>M_S$.
Throughout the viable parameter space $k_\alpha\ll1$ and
$P\to k_\alpha^{2}(g_d^{2}+1)/2g_d$.
Normalizing to the lepton density $n_\ell=(3\zeta_3/2\pi^{2})T^{3}$ per flavor,
Eq.~\eqref{eq:source_master} becomes
\begin{equation}
\dot n_S+3Hn_S=\mathcal{A}(T)\,v_{\rm lep}\,n_\ell(T)\,\mathcal{E}(z)
\sum_\alpha\tanh^{2}\!\big(k_\alpha(T)\big),
\end{equation}
with $z\equiv M_S/T$, so that the entire dependence on the dark mass is
carried by a
single dimensionless efficiency,
\begin{equation}
\mathcal{E}(z)=\frac{2}{3\zeta_3}\int_{z}^{\infty}\!\!du\;
\frac{u^{2}}{e^{u}+1}\,\frac{g_d^{2}+1}{2\,g_d},
\,\, g_d=\sqrt{\frac{u-z}{u+z}} .
\label{eq:Edef}
\end{equation}
With $s=(2\pi^{2}/45)g_{*s}T^{3}$, $H=1.66\sqrt{g_*}T^{2}/M_{\rm pl}$,
$dT/dt=-HT$ and $v_{\rm lep}\approx1$, the yield $Y_S=n_S/s$ follows as
\begin{equation}
\small Y_S=\gamma_{w}\,\hat{\mathcal{C}}\,\frac{m_\sigma^{0}}{T_c^{2}}\,\mathcal{J}
\sum_\alpha|y_\alpha|^{2},
\;
\mathcal{J}\equiv\!\int_{T_{\rm ann}}^{T_{\rm onset}}\!\!
\frac{\mathcal{S}^{2}\,\mathcal{E}}{g_{*s}\sqrt{g_*}}\,\frac{dT}{T},
\label{eq:yield}
\end{equation}
with $\hat{\mathcal{C}}\equiv45\,g_{\rm dof}\zeta_3M_{\rm pl}/2(1.66)\pi^{4}
\simeq3.1\times10^{18}$~GeV.
We keep $g_*(T)$ inside the integral, as it drops from $106.75$ to $\simeq75$
between $T_{\rm onset}$ and $T_{\rm ann}$.

\begin{figure}[t]
\centering
\includegraphics[width=1\linewidth]{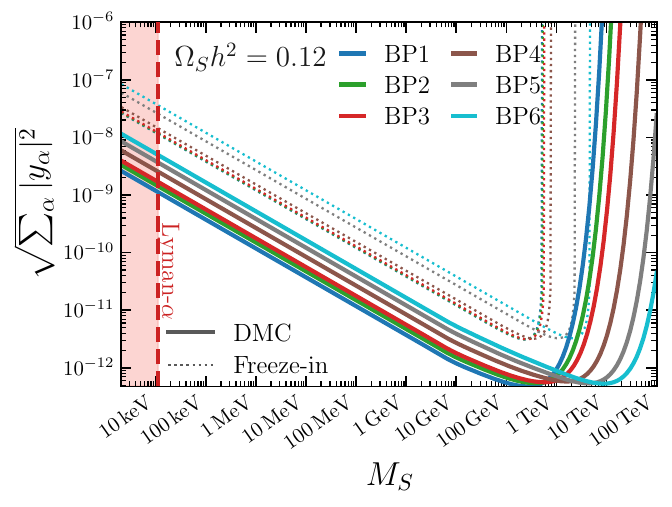}
\caption{Coupling required by $\Omega_S h^2=0.12$ in the
$(M_S,\sum_\alpha|y_\alpha|^2)$ plane for benchmarks BP1--BP6
(Table~\ref{tab:bench}), with $\gamma_w=1$ and $T_{\rm ann}=2$~GeV.
BP1--BP6 are drawn in blue, green, red, brown, gray and cyan, respectively.
Solid: Defect-Mediated Conversion.
Dotted: ordinary freeze-in through the same portal, same color coding
(SM Sec.~\ref{sec:freezein}).
Red band: Lyman-$\alpha$ exclusion.}
\label{fig:fig_mS_sumY}
\end{figure}

The two regimes of the mechanism are the two limits of $\mathcal{E}$.
For $M_S\ll T_{\rm ann}$ the threshold is irrelevant, $\mathcal{E}\to1$
identically, and $\mathcal{J}$ decouples from $M_S$.
Imposing $\Omega_Sh^{2}=0.12$, i.e.\ $M_SY_S=4.4\times10^{-10}$~GeV, then
delivers $M_S$ explicitly,
\begin{equation}
\boxed{\;
\begin{aligned}
M_S&\simeq10~\text{keV}\left(\frac{1}{\gamma_{w}}\right)
\left(\frac{T_c}{400~\text{GeV}}\right)^{2}
\left(\frac{1~\text{TeV}}{m_\sigma^{0}}\right)\\[2pt]
&\times\left(\frac{0.10}{\mathcal{S}^{2}}\right)
\frac{\ln172}{\ln(T_{\rm onset}/T_{\rm ann})}
\left(\frac{1.25\times10^{-9}}
{\sqrt{\textstyle\sum_\alpha|y_\alpha|^{2}}}\right)^{2}\\[2pt]
&\times\frac{g_{*s}\sqrt{g_*}}{(106.75)^{3/2}} .
\end{aligned}\;}
\label{eq:master}
\end{equation}
Eq.~\eqref{eq:master} is obtained by pulling $\mathcal{S}^{2}(T)$ out of
$\mathcal{J}$ at the single value $\mathcal{S}(T_{\rm eval})$, so that
\begin{equation}
\mathcal{J}\;\longrightarrow\;
\frac{\mathcal{S}^{2}(T_{\rm eval})\,\ln(T_{\rm onset}/T_{\rm ann})}
{g_{*s}\sqrt{g_*}} .
\label{eq:Jconst}
\end{equation}
Since $\mathcal{S}$ grows monotonically as the universe cools and
$T_{\rm eval}=0.9\,T_{\rm onset}$ sits near the top of the window, this
constant-portal estimate understates the full $\mathcal{J}$ by a factor
$1.3$ (BP1) to $5.6$ (BP4), the spread reflecting how abruptly the portal
switches on (SM Sec.~\ref{sec:portalnum}).
It is therefore conservative for the yield, $Y_S\propto\mathcal{J}$, but
because $M_S\propto\mathcal{J}^{-1}$ at fixed coupling it overestimates the
dark mass by the same factor: Eq.~\eqref{eq:master} is an analytic guide to
the scalings, while Fig.~\ref{fig:fig_mS_sumY} and
Table~\ref{tab:freezein_compare} use the full numerical $\mathcal{J}$.

The relation $M_S\propto(\sum_\alpha|y_\alpha|^{2})^{-1}$ sweeps the mass axis
from the keV Lyman-$\alpha$ floor upward, and the initial density of the network
enters through $\gamma_w$.
An initially less dense network translates the viable band rigidly
($M_S\propto\gamma_w^{-1}$) without changing its slope, so that saturating
$\gamma_w=1$ gives the minimal Yukawa and the maximal mass reach at fixed wall
scale.
The remaining dependence, through $\mathcal{J}$, is only logarithmic in
$T_{\rm ann}$.
$M_S$ is not a free input but the solution of $\Omega_Sh^{2}=0.12$.

For $M_S\gtrsim T$ the same expressions continue to hold, but $\mathcal{E}$ no
longer factorizes out and Eq.~\eqref{eq:yield} becomes implicit in $M_S$, solved
numerically.
The behavior of $\mathcal{E}$ then controls the high-mass reach, and with it the
shape of the viable band.
Near threshold the fraction of leptons with $E_\ell>M_S$ is still sizable during
the early part of the conversion window, and the probability is enhanced by the
factor $1/g_d$ as $g_d\to0$, so the two effects compensate and $\mathcal{E}$
rises mildly to $\simeq1.12$ at $z\simeq1.5$.
Since $\Omega_Sh^{2}$ fixes $M_SY_S$ and
$\sum_\alpha|y_\alpha|^{2}\propto1/(M_S\mathcal{E})$, a heavier $S$ is paid for
with a smaller Yukawa and the required coupling decreases, until the flux above
threshold is exponentially suppressed,
\begin{equation}
\mathcal{E}(z)\simeq0.70\,z^{5/2}e^{-z},\qquad z\gg1,
\label{eq:Easym}
\end{equation}
the $e^{-M_S/T}$ scarcity of leptons above threshold being softened by the
$z^{5/2}$ threshold enhancement.
$\mathcal{J}$ is dominated by its upper limit $T_{\rm onset}$ once
$M_S>T_{\rm onset}$, so the required coupling grows as $e^{M_S/T_{\rm onset}}$.
The mechanism nonetheless remains operative for $M_S$ well above
$T_{\rm onset}$ and above the mediator mass $m_{h^+}$, at the price of a larger
Yukawa.

Fig.~\ref{fig:fig_mS_sumY} shows the resulting band across both regimes,
together with the coupling required by ordinary freeze-in through the same
portal.
Freeze-in demands a larger Yukawa than DMC everywhere.
Below $M_S=m_{h^+}$ the decay $h^+\to\ell^{+}_\alpha S$ is open and both yields
are linear in $\sum_\alpha|y_\alpha|^{2}$, so the vertical separation between
the two sets of curves is the yield ratio $Y_S/Y_S^{\rm FI}$ of
Table~\ref{tab:freezein_compare}.
At $M_S=m_{h^+}$ that channel closes kinematically, and since each benchmark
carries its own $m_{h^+}$ the freeze-in curves break at different masses.
Beyond the break only $\ell^+\ell^-\to S\bar S$ survives, with a freeze-in yield
proportional to $(\sum_\alpha|y_\alpha|^{2})^{2}$ and requiring two bath
particles above threshold, at a cost $e^{-2M_S/T}$ against the single
exponential of Eq.~\eqref{eq:Easym}.
Therefore, this process is Boltzmann and Yukawa suppressed.
Wall crossing needs only one energetic lepton per conversion, and is the only
one of the two mechanisms that stays viable, at $\sum_\alpha|y_\alpha|^2\ll1$,
for a dark sector heavier than its own mediator.

\textbf{Constraints and signatures} --

\paragraph*{Structure formation and the Lyman-$\alpha$ floor.}
For a light dark sector the momentum distribution of $S$ is inherited from the
bath.
Every $\ell\to S$ event is a local conversion at fixed energy, and because
$P^{\rm rel}_{\ell \rightarrow S}$ is energy-independent in this regime it removes the same
fraction of leptons at every momentum.
The shape of the spectrum is therefore left untouched and $S$ carries the exact
Fermi-Dirac form $f_S(q)=1/(e^{q}+1)$ with $q=p_S/T_S$, independently of when
during the conversion window the event occurred (SM, Sec.~\ref{sec:lyman}).
After the network annihilates, $S$ free-streams as warm dark matter.
Its temperature is fixed by entropy conservation alone,
$\xi_S\equiv T_S/T_\gamma=(3.91/g_{*s})^{1/3}=0.332$, since the wall depletes
the population without cooling it.
Matching the free-streaming length onto the thermal relic of equal $T/m$
(SM, Sec.~\ref{sec:lyman}) gives the equivalent mass
$m_{\rm th}^{\rm eq}=(0.481\,M_S/{\rm keV})^{3/4}$~keV, and imposing
$m_{\rm th}^{\rm eq}>3.5~(5.3)$~keV~\cite{Irsic:2017ixq} excludes
\begin{equation}
M_S<11.0~\text{keV (cons.)},\qquad M_S<19.2~\text{keV (string.)}.
\label{eq:lya_bound}
\end{equation}
This is the only lower bound: heavier $S$ is cold and unconstrained by structure
formation.

\paragraph*{Gravitational waves and colliders.}
Independently of the portal, the annihilating network sources a
gravitational-wave background with
$\Omega_{\rm GW}h^2|_{\rm peak}\propto\sigma_{\rm wall}^2/T_{\rm ann}^4$ and
$f_{\rm peak}\propto T_{\rm ann}$~\cite{Hiramatsu:2010yz,Kawasaki:2011vv,Hiramatsu:2013qaa,Saikawa:2017hiv,Ferreira:2022zzo}.
The annihilation temperature follows from balancing the vacuum energy bias
$\Delta V\simeq2\epsilon v_\sigma^{3}$ against the wall tension,
$T_{\rm ann}\propto\sqrt{\Delta V\,M_{\rm pl}/\sigma_{\rm wall}}$, so at fixed
$(\lambda_\sigma,v_\sigma)$ it is controlled by the $\mathbb{Z}_2$-breaking
coupling $\epsilon$ alone.
Our benchmark $T_{\rm ann}=2$~GeV keeps all three charged leptons thermally
populated in the converting bath and places annihilation before BBN.
The signal is then well below current and projected sensitivities, but
$\Omega_Sh^2$ depends on $T_{\rm ann}$ only logarithmically
[Eq.~\eqref{eq:master}], so smaller $\epsilon$ lowers it at negligible cost in
abundance and moves $f_{\rm peak}$ into the nano-Hz band probed by pulsar timing
arrays and SKA~\cite{NANOGrav:2023gor,NANOGrav:2023hvm}.
Collider phenomenology of the long-lived $h^+$ is given in full in SM
Sec.~\ref{sec:llp_sm}.
Finally, $S$ has no mixing with neutrinos and no coupling to photons or $Z$, so
it produces no X-ray line and no indirect signal at any mass.

\textbf{Conclusion} --
We have shown that cosmological domain walls can drive dark-matter genesis
through a condensate trapped in the wall core.
The crossing probability follows from the exact Dirac solution and depends on
the wall only through the mixing area $k=y_\alpha\mathcal{S}(T)$ and the
kinematics of the produced dark particle.
Imposing $\Omega_Sh^2=0.12$ collapses the parameter space onto
$M_S\propto(\gamma_w\mathcal{S}^{2}\sum_\alpha|y_\alpha|^{2})^{-1}$, a single
band running from the keV Lyman-$\alpha$ floor to the wall-formation scale.
DMC dominates freeze-in through the same portal by $\sim30$--$100$ while the
mediator decay is open, and remains operative even above $m_{h^+}$, where freeze-in
is heavily suppressed.

Nothing in the mechanism requires a minimal dark sector, and the conversion is
not tied to leptons: any fermion coupled to the condensate converts by the same
argument.
It therefore applies to a broad class of $\mathbb{Z}_2$-symmetric extensions in
which a charged condensate forms on the
wall~\cite{Sassi:2023cqp,Battye:2025uny,Law:2021ing,Sassi:2024cyb,Battye:2020sxy},
such as the 2HDM or the N2HDM with dark fermions.
The long-lived $h^\pm$ gives displaced-lepton or stable-charged-track
signatures at the LHC and it is a possible signature of this mechanism.
We quote results for a maximal initial area density ($\gamma_w=1$); a dedicated simulation of the network,
including effects of friction caused by the interaction of the thermal plasma with the domain wall-condensate system, is left to future work.

\textbf{Acknowledgments} --
\begin{acknowledgments}
The authors would like to thank Michael Ramsey-Musolf, Jonas Frerick, Yu Hamada,
Wenyuan Ai, Vinicius Oliveira, Sk Jeesun and Jacinto P. Neto for useful
discussions.
JPP is supported by the National Natural Science Foundation of China
(12425506 and 12375101).
\end{acknowledgments}

\addcontentsline{toc}{section}{References}
\bibliographystyle{apsrev4-2}
\bibliography{apssamp.bib}

\clearpage
\onecolumngrid
\setcounter{secnumdepth}{3}
\setcounter{section}{0}
\renewcommand{\thesection}{\Roman{section}}
\renewcommand{\theequation}{S\arabic{equation}}
\setcounter{equation}{0}

\begin{center}
{\large\bfseries Supplemental Material}\\[4pt]
{\normalsize Defect-Mediated Conversion: Dark Matter from Cosmological Domain-Wall Scattering}
\end{center}
\vspace{6pt}

\section{Scalar field condensate inside the wall}
\label{sec:dw}

That a field can acquire a vacuum expectation value confined to the core of a
topological defect while vanishing outside it is a long-known
phenomenon~\cite{Witten:1984eb}, originally identified for charged condensates
on cosmic strings.
In this section we discuss in detail the emergence of a condensate for the field
$h^+$ in the core of the domain walls. 

The prerequisite is a negative effective mass term for $h^+$ inside the core,
leading to spontaneous symmetry breaking in a region localized inside and in the
vicinity of the wall.
In the background of a domain wall profile $\sigma(x)$, this effective mass is
\begin{equation}
M^2_+ (x) = \dfrac{1}{2} \mu^2_{+} + \lambda_{\sigma +} \sigma^2(x).
\end{equation}
Far from the wall, $M^2_{+}=\tfrac{1}{2}\mu^2_{+}+\lambda_{\sigma+}v^2_\sigma>0$,
leading to a vanishing VEV for $h^+$.
Inside the wall, $M^2_+(0)=\tfrac{1}{2}\mu^2_{+}$, which can be negative.
In that case the scalar potential in the $h^+$ direction can develop a non-zero
minimum.
A non-zero condensate $\langle h^+(0)\rangle$ nevertheless contributes to the
energy density of the field configuration through the kinetic term
$\tfrac{1}{2}(dh_+/dx)^2$.
Determining whether a condensate forms therefore requires solving the static
equations of motion for the scalar fields with the domain wall boundary
conditions,
\begin{align}
& \dfrac{1}{2} \dfrac{d^2\phi}{dx^2} - \biggl[ \dfrac{1}{2}\mu^2_{\rm sm} \phi(x) + \dfrac{1}{4} \lambda_{\rm sm} \phi^3(x) + v(x)\biggl( \lambda_{\Phi\sigma} \sigma^2(x) + \lambda_{\Phi+}h^2_+(x) \biggr) \biggr] =0
\\
& \dfrac{d^2\sigma}{dx^2} - \biggl[ \mu^2_{\sigma} \sigma(x) + \lambda_{\sigma} \sigma^3(x) + 3 \epsilon \sigma^2(x) + \sigma(x)\biggl( 2\lambda_{\sigma+} h^2_+(x) + \lambda_{\Phi\sigma}\phi^2(x) \biggr) \biggr] = 0 \\
& \dfrac{d^2h_{+}}{dx^2} - \biggl[ \mu^2_{+} h_+(x) + \lambda_{+} h_+^3(x) + h_{+}(x)\biggl( 2\lambda_{\sigma+} \sigma^2(x) + \lambda_{\Phi+}\phi^2(x) \biggr) \biggr] =0.
\end{align}

\begin{figure}
\centering
\includegraphics[width=1\linewidth]{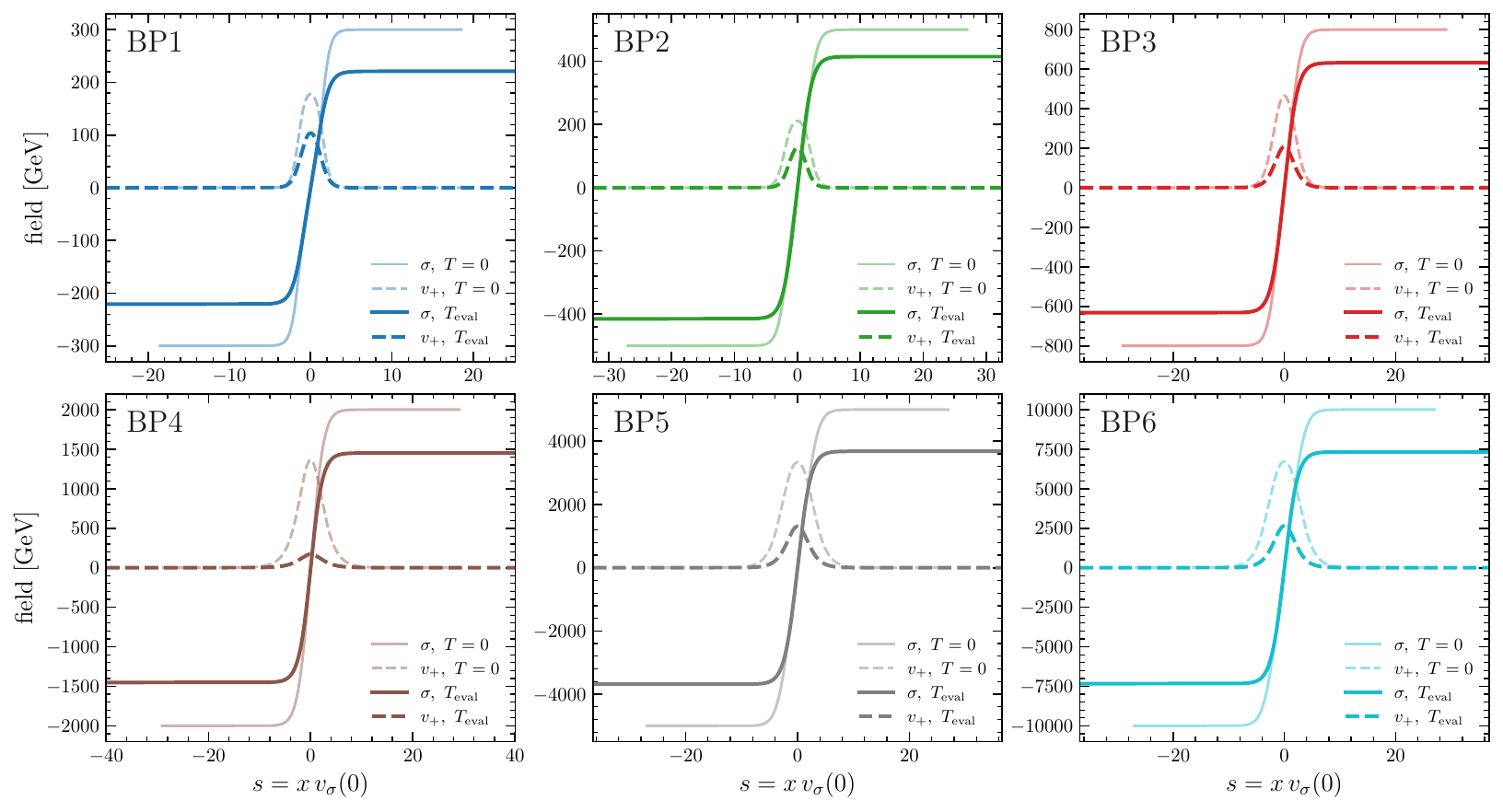}
\caption{Field profiles $\sigma(x)$ (solid) and $v_+(x)$ (dashed) for benchmarks
BP1--BP6, shown both at $T=0$ (light, thin) and at $T_{\rm eval}$ (bold), as a
function of $s=x\,v_\sigma(0)$.}
\label{fig:dw_bps_profile}
\end{figure}

The solution determines the profile of the scalar fields by minimizing the
energy of the configuration.
Since this system is non-trivial, we solve it numerically following the
gradient-flow method introduced in~\cite{Battye:2011jj} in the context of domain
wall solutions.
The resulting profiles for the benchmark points are shown in
Fig.~\ref{fig:dw_bps_profile}.

To determine the range of parameters leading to a condensate
$\langle h^+\rangle=v_+$ inside the wall, one can study the stability of the
solution $h_+(x)=0$ everywhere.
Sizable couplings $\lambda_{\Phi+}$ and $\lambda_{\Phi\sigma}$ can lead to
significant back-reaction of the SM field profile on the wall and condensate
solutions, so for simplicity we treat the case in which the coupling between the
SM-like scalar and the extra scalars is very small or zero.

We follow the approach of~\cite{Vachaspati:2006zz}.
Consider a small time-dependent perturbation $\delta_+(x)\cos(\omega t)$ around
the solution $h_+(x)=0$.
If that solution is not the lowest-energy one, then $\omega^2<0$ and the
perturbation grows with time, signaling the instability.
In the background of the domain wall profile
$\sigma(x)=v_\sigma\tanh(x/\delta_{\rm wall})$, the linearized equation of
motion for $\delta_+$ is
\begin{equation}
-\dfrac{d^2\delta_+}{dx^2} + \biggl[\mu^2_+ + 2\lambda_{\sigma+} v^2_\sigma \tanh^2\Big(\dfrac{x}{\delta_{\rm wall}}\Big)  \biggr] \delta_+(x) = \omega^2 \delta_+(x).
\label{eq:stabilityv+}
\end{equation}
This is a Schrödinger-type equation with $\omega^2$ identified as the energy of
an eigenstate $\delta_+$ in the potential
$U(x)=\mu^2_++2\lambda_{\sigma+}v^2_\sigma\tanh^2(x/\delta_{\rm wall})$, a well
with positive values far from the wall and possibly negative values inside it,
admitting bound states that act as a condensate for $h_+(x)$.
We seek the parameters for which $h_+(x)=0$ is unstable, i.e.\ bound-state
fluctuations with negative eigenvalue $\omega^2$.
Eq.~\eqref{eq:stabilityv+} can be rewritten as~\cite{Vachaspati:2006zz}
\begin{equation}
-\dfrac{d^2\delta_+}{dx^2} + \dfrac{2\lambda_{\sigma+}v^2_\sigma}{3}\biggl[3 \tanh^2\Big(\dfrac{x}{\delta_{\rm wall}}\Big) - 1\biggr] \delta_+(x) = \Big(\omega^2  - \dfrac{2\lambda_{\sigma+}v^2_\sigma}{3}- \mu^2_+ \Big) \delta_+(x) = \hat{\omega}^2 \delta_+(x),
\end{equation}
where $\hat{\omega}^2=\omega^2-\tfrac{2}{3}\lambda_{\sigma+}v^2_\sigma-\mu^2_+$.
This equation has a lowest-energy bound state at $\hat{\omega}^2=0$ with
$\delta_+(x)=\sech^2(x)$~\cite{Vachaspati:2006zz}.
Therefore, for $\mu^2_+<-\tfrac{2}{3}\lambda_{\sigma+}v^2_\sigma$ the eigenvalue
$\omega^2$ is negative, the solution $h_+(x)=0$ is unstable, and a non-vanishing
condensate $\langle h^+\rangle(x)$ develops inside the wall.
Since $\mu^2_+=2m^2_{h^+}-2\lambda_{\sigma+}v^2_{\sigma}$, this translates into
an upper bound on the charged scalar mass,
\begin{equation}
m^2_{h^+} < \dfrac{2}{3}\lambda_{\sigma+}v^2_\sigma .
\end{equation}

A second constraint follows from vacuum decay induced by domain
walls~\cite{Sassi:2025dyj}.
Consider a potential with two coexisting minima: the asymptotic one far from the
wall, $(\langle\sigma\rangle=\pm v_\sigma,\langle h^+\rangle=0)$, denoted
$\mathcal{N}$, and another with $(\langle\sigma\rangle=0,\langle h^+\rangle=c_+)$,
denoted $\mathcal{CB}$.
These are separated by a barrier that prevents tunneling.
When $\mathcal{N}$ is the global minimum, a condensate corresponding to
$\mathcal{CB}$ is stable and bounded inside the wall.
If instead $\mathcal{CB}$ is the global minimum, it is nucleated inside the wall
and expands outside it, populating the whole universe with the $\mathcal{CB}$
vacuum and destroying the network.

Such a scenario can be desirable as an intermediate step before an electroweak
SM-like minimum with $v_{\rm sm}=246$~GeV becomes the global minimum, and
constitutes a natural way to annihilate the network without a
$\mathbb{Z}_2$-breaking term.
For simplicity, and to focus on the case where the condensate is stable and
localized inside the wall, we take parameters for which $\mathcal{N}$ lies below
$\mathcal{CB}$.
Following~\cite{Sassi:2025dyj}, the difference between the two minima is
\begin{equation}
V_{\mathcal{N}} - V_{\mathcal{CB}} = \dfrac{1}{2}\biggl[ v^2_\sigma \biggl( \dfrac{1}{2}\mu^2_\sigma + \lambda_{\sigma +}c^2_+ \biggr) - m^2_{h^+} c^2_+ \biggr] = \dfrac{1}{2} \biggl[ \mu^2_\sigma v^2_\sigma - \mu^2_+ c^2_+ \biggr].
\end{equation}
Using the minimization conditions $\mu^2_++\lambda_+c^2_+=0$ and
$\mu^2_\sigma+\lambda_\sigma v^2_\sigma=0$,
\begin{equation}
V_{\mathcal{N}} - V_{\mathcal{CB}} = \dfrac{1}{4}\biggl[ \dfrac{\mu^4_+}{\lambda_+} - \dfrac{\mu^4_\sigma}{\lambda_\sigma}  \biggr],
\end{equation}
so the condition for a stable bounded condensate is
\begin{equation}
\dfrac{\mu^4_+}{\lambda_+} < \dfrac{\mu^4_\sigma}{\lambda_\sigma},
\end{equation}
which can be expressed as a lower bound on $m_{h^+}$,
\begin{equation}
m^2_{h^+} > \dfrac{1}{2}\biggl( 2\lambda_{\sigma+}v^2_{\sigma} - \sqrt{\lambda_+} v_\sigma m_{\sigma} \biggr).
\end{equation}
Combining both constraints gives the window in which the mechanism operates,
\begin{equation}
\dfrac{1}{2}\biggl( 2\lambda_{\sigma+}v^2_{\sigma} - \sqrt{\lambda_+} v_\sigma m_{\sigma} \biggr) < m^2_{h^+} < \dfrac{2}{3}\lambda_{\sigma+}v^2_{\sigma}.
\end{equation}

\section{Temperature dependent effects on the scalar potential}
\label{sec:thermal}

We here discuss the effects of the LO thermal corrections to the scalar
potential on the profiles of the domain walls and of the condensate.

Taking the simplified case in which the SM Higgs couplings to the BSM scalars
are negligible, the leading-order thermal corrections to the quadratic terms for
$\sigma$ and $h^+$ are~\cite{Dolan:1973qd,Quiros:1999jp}
\begin{align}
\mu^2_+(T) &= \mu^2_+ + \dfrac{1}{12}(3\lambda_+ + 2 \lambda_{\sigma+} )T^2, \\
\mu^2_\sigma(T) &= \mu^2_\sigma + \dfrac{1}{12}(3\lambda_\sigma + 2 \lambda_{\sigma+} )T^2.
\end{align}
Since $\lambda_{\sigma+}>0$ by definition in order to obtain a condensate inside
the wall, the thermal masses $\mu^2_+(T)$ and $\mu^2_\sigma(T)$ become positive
above the temperatures $T^c_{+}$ and $T^c_{\sigma}$, respectively.

\begin{figure}
\centering
\includegraphics[width=1\linewidth]{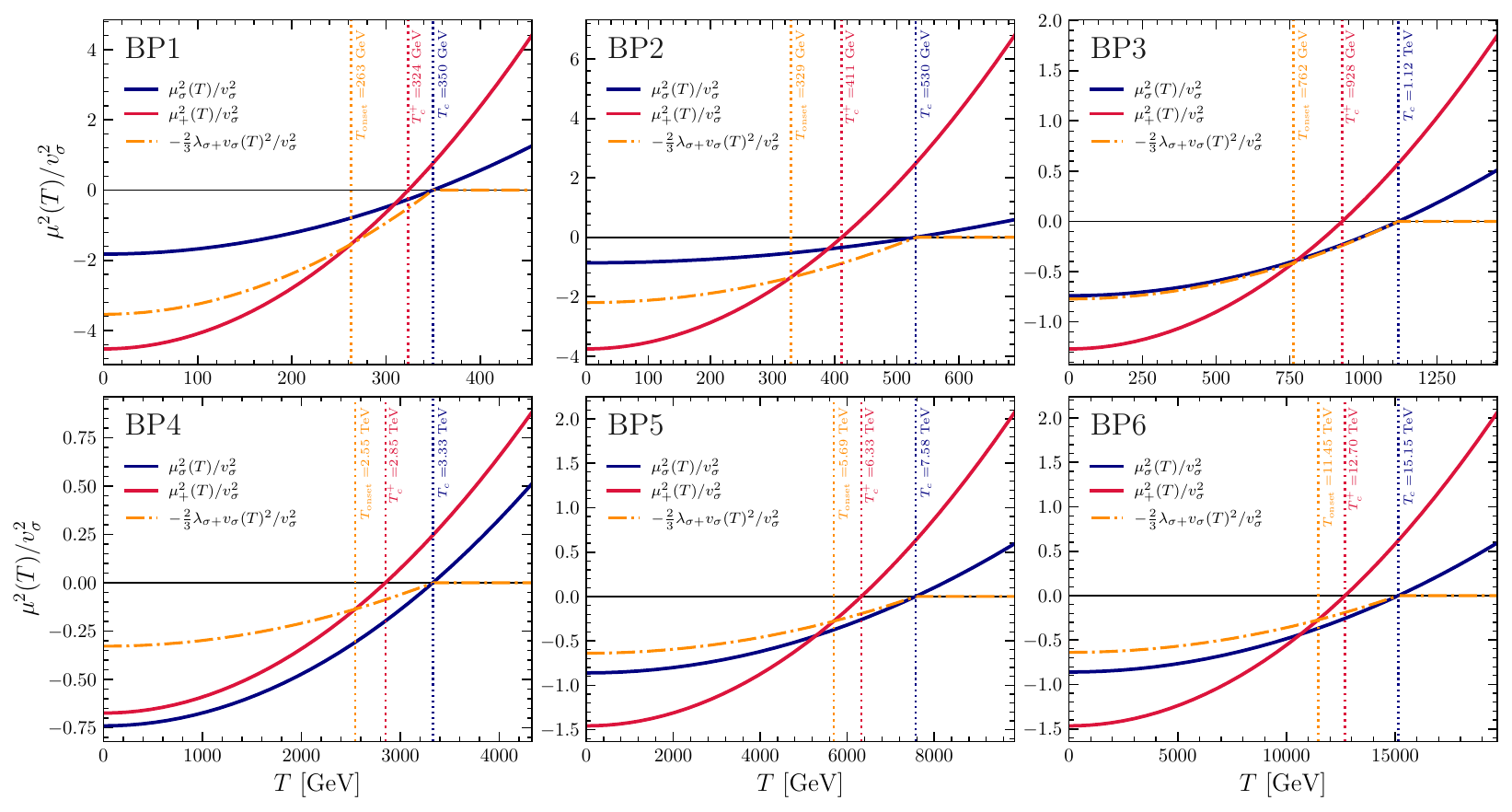}
\caption{Thermal evolution of $\mu_\sigma^2(T)/v_\sigma^2$ (blue) and
$\mu_+^2(T)/v_\sigma^2$ (red), together with the threshold curve
$-\tfrac23\lambda_{\sigma+}v_\sigma(T)^2/v_\sigma^2$ (orange, dash-dotted), for
benchmarks BP1--BP6.
Vertical dotted lines mark $T_c$, $T_{c}^+$ and $T_{\rm onset}$ for each panel,
with labels color-matched to the corresponding curve.}
\label{fig:dw_bps_thermal}
\end{figure}

For $T<T^c_\sigma$, $\sigma$ acquires a non-zero VEV and domain walls form.
At that point $\mu^2_+(T)$ can still be positive, so no $\langle h^+\rangle$
condensate develops inside the wall.
Below $T^c_+$, $\mu^2_+(T)$ turns negative, but the solution $v_+(x)=0$
everywhere remains stable.
The instability leading to a non-zero condensate is reached only when
$\mu^2_+(T_{\rm onset})<-\tfrac{2}{3}\lambda_{\sigma+}v^2_\sigma(T_{\rm onset})$
(see Fig.~\ref{fig:dw_bps_thermal}).

\section{Scaling of the mixing area and its numerical treatment}
\label{sec:portalnum}

\subsection{Cancellation of $v_\sigma$}
For thin walls, the mixing area of Eq.~\eqref{eq:Sdef} is fixed by the
dimensionless quartics alone.
The in-core condensate $v_+(0)$ grows with both $v_\sigma$ and
$\lambda_{\sigma+}$, while the wall width $2/m_\sigma$ shrinks with $v_\sigma$
and $\lambda_\sigma$, so the two dependences cancel in the product and
\begin{equation}
\mathcal{S}=\frac{2v_+(0)}{m_\sigma}
=\mathcal{S}(\lambda_\sigma,\lambda_{\sigma+},\lambda_+),
\end{equation}
independently of the symmetry-breaking scale.
The range of $\mathcal{S}$ accessible to the mechanism is therefore bounded from
above by the tree-level perturbativity requirement
$\lambda_i\lesssim4\pi$~\cite{Lee:1977eg}, which BP1 and BP2 saturate, rather
than by any dimensionful parameter.

This cancellation is verified numerically by BP5 and BP6, which differ by a
factor of two in $v_\sigma$ (5 to 10~TeV) at fixed quartics
$(\lambda_\sigma,\lambda_{\sigma+},\lambda_+)=(0.86,0.96,3.0)$, yet whose
$\mathcal{S}(0)$ agree to better than $1\%$: $1.018$ and $1.025$ respectively.
It is what allows the mechanism to be moved up and down the mass axis without
retuning the portal, and it is the reason why $\gamma_w$, $m_\sigma^0$ and $T_c$
in Eq.~\eqref{eq:master} carry the entire scale dependence.

\subsection{The thin-wall condition}
The collapse of Eq.~\eqref{eq:Pfull} to $\tanh^2(k)$ requires the thin-wall
regime $\delta_{\rm wall}\ll\lambda_{\rm dB}$, in which the de Broglie
wavelength $\lambda_{\rm dB}=2\pi/p$ of the incident lepton cannot resolve the
wall profile and the condensate acts as a delta-function insertion.
For thick walls the scattering must instead be treated in the WKB regime, where
$\lambda_{\rm dB}$ re-enters and the transition probability acquires a
form-factor suppression depending on the shape of the profile; the benchmarks of
Table~\ref{tab:bench} are all chosen to satisfy the thin-wall requirement, with
wall width below $5/v_\sigma$ both at $T=0$ and at $T_{\rm eval}$.

\subsection{Numerical treatment of the portal}
We do not freeze the portal at a single temperature.
The coupled scalar equations of motion are solved at each $T$ across the
conversion window, giving $\mathcal{S}(T)$, which is then integrated numerically
in Eq.~\eqref{eq:yield} together with the running $g_*(T)$ and the threshold
efficiency $\mathcal{E}(M_S/T)$.

The constant-portal estimate quoted in the main text as an analytic cross-check
replaces $\mathcal{S}^{2}(T)$ by its value at
$T_{\rm eval}=0.9\,T_{\rm onset}$ throughout the window.
Since $\mathcal{S}$ grows monotonically as the universe cools, this is
conservative: it understates the full $\mathcal{J}$ by a factor $1.3$ (BP1) to
$5.6$ (BP4), with the spread reflecting how close $T_{\rm onset}$ sits to
$T_c^{+}$, the temperature at which $\mu_+^{2}(T)$ turns negative.
Benchmarks for which the two temperatures are nearly degenerate have a portal
that switches on abruptly and then grows substantially over the window, so the
constant-portal estimate is worst there.

\section{Wall domination and the cosmological history}
\label{sec:walldom}

The energy density of a wall network redshifts as $\rho_{\rm dw}\propto a^{-1}$,
more slowly than both matter ($a^{-3}$) and radiation ($a^{-4}$), so an
unbiased network would eventually come to dominate the expansion.
This occurs at a time~\cite{Press:1989yh,Vilenkin:2000jqa}
\begin{equation}
t_{\rm dom}=\frac{3M_{\rm pl}^{2}}{4\,\mathcal{A}_{\rm sc}\,\sigma_{\rm wall}}
\approx2.93\times10^{3}~{\rm s}\;\mathcal{A}_{\rm sc}^{-1}
\left(\frac{\sigma_{\rm wall}}{{\rm TeV}^{3}}\right)^{-1},
\label{eq:tdom}
\end{equation}
with $\mathcal{A}_{\rm sc}\simeq0.8$ the area parameter of the scaling solution
measured in lattice simulations~\cite{Hiramatsu:2013qaa}, not to be confused
with the area density $\mathcal{A}(T)$ of Eq.~\eqref{eq:area}, and
$\sigma_{\rm wall}=\tfrac{2\sqrt2}{3}\sqrt{\lambda_\sigma}\,v_\sigma^{3}$ the
wall tension.

Since $\sigma_{\rm wall}\propto\sqrt{\lambda_\sigma}\,v_\sigma^{3}$, a network
that survived until $T=1$~GeV would dominate the universe only for
$v_\sigma\sim10^{7}$~GeV, three orders of magnitude above our heaviest
benchmark.
For BP1--BP6 the bias term annihilates the network at $T_{\rm ann}$ long before
$t_{\rm dom}$, so the radiation-dominated history assumed throughout, in
particular in Eq.~\eqref{eq:yield} and in the redshifting of
Eq.~\eqref{eq:area}, holds over the entire conversion window.
The mechanism is therefore insensitive to the details of wall domination, and
the only role of the $\mathbb{Z}_2$-breaking coupling $\epsilon$ is to set
$T_{\rm ann}$, on which the relic abundance depends only logarithmically
[Eq.~\eqref{eq:master}].

\begin{table}[h]
\centering
\caption{Full parameter set for benchmarks BP1--BP6, all satisfying the
thin-wall requirement (wall width $<5/v_\sigma$) both at $T=0$ and at
$T_{\rm eval}=0.9\,T_{\rm onset}$.
The in-core condensate $v_+$ is obtained by solving the coupled scalar equations
of motion at each temperature.
Note that BP5 and BP6 share the same quartics and differ only in $v_\sigma$,
yielding the same $\mathcal{S}$ to better than $1\%$.
Masses in GeV.}
\label{tab:bench_full}
\begin{ruledtabular}
\begin{tabular}{lccccccccccc}
& $v_\sigma$ & $m_{h^+}$ & $\lambda_\sigma$ & $\lambda_{\sigma+}$ & $\lambda_{+}$ & $v_+(0)$ & $v_+(T_{\rm eval})$ & $T_c$ & $T_{\rm onset}$ & $T_{\rm eval}$ & $\mathcal{S}(0)$ \\
\hline
BP1 & 300   & 523  & 1.82 & 5.31 & 12.0 & 172  & 104  & 350   & 263   & 237   & 0.62 \\
BP2 & 500   & 542  & 1.16 & 2.26 & 5.0  & 276  & 152  & 660   & 463   & 417   & 0.78 \\
BP3 & 800   & 579  & 0.74 & 1.16 & 3.0  & 388  & 208  & 1119  & 762   & 685   & 0.90 \\
BP4 & 2000  & 789  & 0.74 & 0.49 & 1.0  & 638  & 167  & 3329  & 2545  & 2291  & 1.12 \\
BP5 & 5000  & 2399 & 0.86 & 0.96 & 3.0  & 3021 & 1308 & 7575  & 5691  & 5122  & 1.018 \\
BP6 & 10000 & 4747 & 0.86 & 0.96 & 3.0  & 6086 & 2643 & 15150 & 11449 & 10304 & 1.025 \\
\end{tabular}
\end{ruledtabular}
\end{table}
\section{Dirac equation and conversion probability}
\label{sec:dirac}
The one-dimensional Dirac problem in a wall background is standard, both for
fermion modes bound to the defect~\cite{Jackiw:1975fn} and for the reflection
and transmission of bath fermions used in electroweak
baryogenesis~\cite{Cohen:1993nk,Joyce:1994zn,Joyce:1994zt}.
We adapt it here to a two-species system in which the off-diagonal mass is
supplied by the in-core condensate.

To describe the scattering of SM thermal bath fermions off the wall and their
transformation into DM particles, we solve the one-dimensional Dirac equations
for the SM fermions and the DM particles in the background of the condensate
$v_+(x)$.

The relevant Lagrangian is
\begin{equation}
\mathcal{L}  \supset i\overline{\ell_\alpha}\slashed{\partial}\ell_\alpha  + i\bar{S}\slashed{\partial}S - m_{\ell,\alpha}(x)\overline{\ell_\alpha}\ell_\alpha - M_S\,\bar S S -y_\alpha\, v_+(x) \,\overline{\ell^{R}_{\alpha}}\,S
+\mathrm{h.c.},
\end{equation}
leading to the system of Dirac equations for the SM particle (here a charged
lepton $\ell_\alpha$) and the dark fermion $S$,
\begin{equation}
i\slashed{\partial} \begin{pmatrix}
\ell_\alpha \\
S
\end{pmatrix} - \begin{pmatrix}
m_{\ell,\alpha}(x) && y_\alpha v_+(x) P_L \\
y_\alpha v_+(x) P_R && M_S
\end{pmatrix} \begin{pmatrix}
\ell_\alpha \\
S
\end{pmatrix} = 0,
\label{eq:systemdirac}
\end{equation}
where $P_R$ and $P_L$ are the right- and left-handed projectors.
We take plane-wave solutions describing the scattering states of the incoming,
reflected and transmitted particles,
\begin{align}
\ell_\alpha (x,t) &= e^{-iE_\ell t} \ell^{\rm inc}_\alpha(x) +  e^{-iE_\ell t} \ell^{\rm ref}_\alpha(x) \text{ for }x<0, \\
\ell_\alpha (x,t) &= e^{-iE_\ell t} \ell^{\rm tra}_\alpha(x) \text{ for }x>0, \\
S (x,t) &= e^{-iE_\ell t} S^{\rm ref}(x) \text{ for }x<0, \\
S (x,t) &= e^{-iE_\ell t} S^{\rm tra}(x)\text{ for }x>0,
\end{align}
where $\ell^{\rm inc}_{\alpha}$ denotes the incoming charged lepton spinor,
$\ell^{\rm ref}_{\alpha}$ the reflected spinor and $\ell^{\rm tra}_{\alpha}$ the
transmitted spinor, with the same notation for the dark fermion $S$.

To obtain analytical solutions we use the thin-wall approximation for the
profile of the wall and of the $v_+(x)$ condensate, expressing
\begin{equation}
y_\alpha v_+(x) = k \, \delta (x),
\end{equation}
where $k=\int dx\, y_\alpha v_+(x)\approx y_\alpha v_+(0)\,\delta_{\rm wall}$,
consistent with Eq.~\eqref{eq:Sdef} of the main text.

Outside the wall the solutions are the standard free spinors.
Matching at $x=0$ by integrating Eq.~\eqref{eq:systemdirac} between $-\epsilon$
and $\epsilon$,
\begin{equation}
\begin{pmatrix}
\ell_\alpha(0^+) \\
S(0^+)
\end{pmatrix} = \hat{P} \exp \biggl[ \int^{\epsilon}_{-\epsilon} dx \begin{pmatrix}
m_{\ell,\alpha}(x) && y_\alpha v_+(x) P_L \\
y_\alpha v_+(x) P_R && M_S
\end{pmatrix} \biggr] \begin{pmatrix}
\ell_\alpha(0^-) \\
S(0^-)
\end{pmatrix} ,
\label{eq:systemdirac2}
\end{equation}
and the result of this exponential integral is~\cite{Sassi:2023cqp}
\begin{equation}
\begin{pmatrix}
\ell_\alpha(0^+) \\
S(0^+)
\end{pmatrix} =  \begin{pmatrix}
\cosh^2(\frac{k}{2}) \,\mathbb{I} + \sinh^2(\frac{k}{2}) \gamma_5  && -i \sinh(k) \gamma_1 P_L  \\
-i \sinh(k) \gamma_1 P_R && \cosh^2(\frac{k}{2}) \,\mathbb{I} - \sinh^2(\frac{k}{2}) \gamma_5
\end{pmatrix} \begin{pmatrix}
\ell_\alpha(0^-) \\
S(0^-)
\end{pmatrix} .
\label{eq:systemdirac_2}
\end{equation}
The reflected and transmitted spinors follow from
\begin{align}
\ell_\alpha(0^-) = \ell^{\rm inc}_\alpha + \ell^{\rm ref}_\alpha, && \ell_\alpha(0^+) = \ell^{\rm tra}_\alpha, &&
S(0^-) = S^{\rm ref}, && S(0^+) = S^{\rm tra},
\end{align}
and the reflection and transmission coefficients are
\begin{align}
T_S(p) &= \dfrac{\mathcal{F}_S^{\rm tra}}{\mathcal{F}^{\rm inc}} =  \dfrac{\overline{S^{\rm tra}}\gamma_1 S^{\rm tra}}{p}\\
R_S(p) &= \dfrac{\mathcal{F}_S^{\rm ref}}{\mathcal{F}^{\rm inc}} = - \dfrac{\overline{S^{\rm ref}}\gamma_1 S^{\rm ref}}{p}\\
T_\ell(p) &= \dfrac{\mathcal{F}_\ell^{\rm tra}}{\mathcal{F}^{\rm inc}} =   \dfrac{\overline{\ell_{\alpha}^{\rm tra}}\gamma_1 \ell_{\alpha}^{\rm tra}}{p}\\
R_\ell(p) &= \dfrac{\mathcal{F}_\ell^{\rm ref}}{\mathcal{F}^{\rm inc}} = - \dfrac{\overline{\ell_{\alpha}^{\rm ref}}\gamma_1 \ell_{\alpha}^{\rm ref}}{p},
\end{align}
with $\mathcal{F}$ denoting the probability current and $p$ the momentum of the
incoming SM particle.
In the limit of small SM fermion mass, the reflection and transmission rates
into an $S$ dark matter particle are
\begin{align}
R_S(p) &= \dfrac{4g_d(g_d+1)^2\sinh^2(k)}{[(g_d-1)^2 - (g_d+1)^2\cosh(k)]^2}, \\
T_S(p) &= \dfrac{4g_d(g_d-1)^2\sinh^2(k)}{[(g_d-1)^2 - (g_d+1)^2\cosh(k)]^2},
\end{align}
where $g_d=\sqrt{E_\ell^2-M^2_S}/(E_\ell+M_S)$.
The total conversion probability of a SM thermal bath fermion into an $S$
particle is therefore
\begin{equation}
P_{\ell_\alpha \rightarrow S} = \dfrac{8g_d(g^2_d + 1)\sinh^2(k)}{[(g_d-1)^2 - (g_d+1)^2\cosh(k)]^2}.
\end{equation}
The wall enters only through the dimensionless $k_\alpha$, and the kinematics
only through the single threshold variable $g_d$, which runs from $g_d=0$ at
$E_\ell=M_S$ to $g_d\to1$ for $E_\ell\gg M_S$.

We distinguish two limits for the produced dark sector particle.
The relativistic limit, $E_\ell\gg M_S$, in which essentially all thermal bath
leptons can be converted, and the near-threshold regime, $E_\ell\gtrsim M_S$, in
which part of the bath does not carry enough energy to be converted as it hits
the wall.
In the relativistic limit $g_d\to1$ and the conversion probability simplifies to
\begin{equation}
P^{\rm rel}_{\ell \rightarrow S} =  \tanh^2(k),
\end{equation}
which is the limit used in Eq.~\eqref{eq:Pgeneric}.
Near threshold, for small $k$, it simplifies instead to
\begin{equation}
P^{\rm thr}_{\ell \rightarrow S} \approx \dfrac{8g_d\, k^2}{(4g_d + k^2/2)^2},
\end{equation}
which for $E_\ell\approx M_S$ and $k^2=8g_d$ approaches
$P^{\rm thr}_{\ell\rightarrow S}\approx1$, a threshold enhancement of the
conversion probability.

\section{Free-streaming and the Lyman-$\alpha$ floor}
\label{sec:lyman}

$S$ does not acquire its own, model-dependent phase-space distribution.
If $S$ is relativistic, it inherits the distribution of the charged lepton $\ell$ it converts
from, because every $\ell\to S$ scattering event off the domain wall is a local,
energy-conserving relabeling of the same on-shell particle.
The conversion probability $P^{\rm rel}_{\ell \rightarrow S}=\tanh^2(k)$, with
$k=y_\alpha\mathcal{S}$, depends only on the wall parameters and not on the
momentum $p$ of the incoming lepton, so the conversion removes a fixed fraction
$P^{\rm rel}_{\ell \rightarrow S}$ of leptons at every $p$ and reinstates them as $S$ with the
same $p$.
The bath lepton at temperature $T$ has $f_\ell(p,T)=1/(e^{p/T}+1)$.
In the relativistic regime the physical momentum of a freely streaming particle
scales as $p(T)=p_0\,T/T_{\rm ann}$, with the comoving momentum
$p_{\rm com}=p_0$ constant.
Therefore
\begin{equation}
f_S\!\left(p,T_{\rm ann}\right)
= P^{\rm rel}_{\ell \rightarrow S}\,f_\ell\!\left(p,T_{\rm ann}\right)
= \frac{P^{\rm rel}_{\ell \rightarrow S}}{e^{p/T_{\rm ann}}+1},
\end{equation}
and this expression is independent of when production occurred at fixed
$g_{*s}$: contributions from any $T\in[T_{\rm ann},T_{\rm onset}]$ all give the
same Fermi-Dirac form evaluated at $T_{\rm ann}$.
Integrating over the production epoch introduces no shape distortion.
Because $g_{*s}$ falls from $106.75$ to $75.75$ across the window, particles
produced late are warmer by up to $12\%$ in $\xi_S\equiv T_S/T_\gamma=(3.91/g_{*s})^{1/3}$, so the exact spectrum is a
superposition of Fermi-Dirac forms with $\xi_S\in[0.332,0.372]$.
We evaluate the bound at the colder endpoint throughout, which is conservative.
In terms of the dimensionless comoving momentum $q\equiv p_S/T_S$, the
distribution is
\begin{equation}
f_S(q) = \frac{1}{e^q+1},
\label{eq:fS_exact}
\end{equation}
exact for $M_S\ll T_{\rm ann}$, with threshold corrections
$\mathcal{O}(M_S^2/T_{\rm ann}^2)$.

Because $f_S$ is a Fermi-Dirac form, the free-streaming length of $S$ can be
mapped exactly onto that of a thermal relic, the two populations differing only
in temperature and normalization.
The mapping is fixed by noting that the reference relic has no free
temperature: it is set by its own mass through
$\Omega_X h^2=(T_X/T_\nu)^3(m_X/93.14~\text{eV})$, so a heavier reference relic
is necessarily colder.
Equating $T/m$ between the two,
\begin{equation}
M_S = \frac{\xi_S\;m_{\rm th}^{4/3}}
{0.714\,\big(0.12\times93.14~\text{eV}\big)^{1/3}}
\;\Longleftrightarrow\;
m_{\rm th}^{\rm eq}=\left(0.481\,\frac{M_S}{\rm keV}\right)^{3/4}\text{keV},
\label{eq:mth_eq}
\end{equation}
evaluated at $\xi_S=0.332$.
The conversion probability suppresses the abundance but
not the spectral shape, which is what free-streaming responds to.

Applying the Lyman-$\alpha$ limits $m_{\rm th}>3.5~(5.3)$~keV of
Ref.~\cite{Irsic:2017ixq} to Eq.~\eqref{eq:mth_eq} gives
\begin{equation}
M_S > 11.0~\text{keV (cons.)}, \qquad M_S > 19.2~\text{keV (string.)}.
\end{equation}

The same conclusion follows from the half-mode
criterion~\cite{Bode:2000gq,Viel:2005qj}, which we use as a cross-check.
The squared transfer function $T^2(k)=P_{\rm WDM}(k)/P_{\rm CDM}(k)$ is fit to
\begin{equation}
T^2(k) = \left[1+(\alpha k)^{2\nu}\right]^{-10/\nu},
\qquad \nu=1.12,
\label{eq:T2fit}
\end{equation}
with $\alpha(m_{\rm th})$ the standard fit of Ref.~\cite{Viel:2005qj}, and the
half-mode defined by $T^2(k_{1/2})\equiv1/2$,
\begin{equation}
k_{1/2} = \frac{1}{\alpha}\left[2^{\nu/10}-1\right]^{1/(2\nu)}.
\label{eq:khalf_supp}
\end{equation}
A model is excluded if $k_{1/2}<k_{1/2}^{\rm WDM}$, its small-scale cutoff
falling at scales too large to be consistent with the observed forest power.
We evaluate both $k_{1/2}(M_S)$, through Eq.~\eqref{eq:mth_eq}.
This yields $k_{1/2}^{\rm WDM}=26.7$ and $42.3\,h/$Mpc for the two reference
masses, some $20\%$ below the values quoted in Ref.~\cite{Irsic:2017ixq} from
their own calibration.

\begin{figure}[h]
\centering
\includegraphics[width=0.7\linewidth]{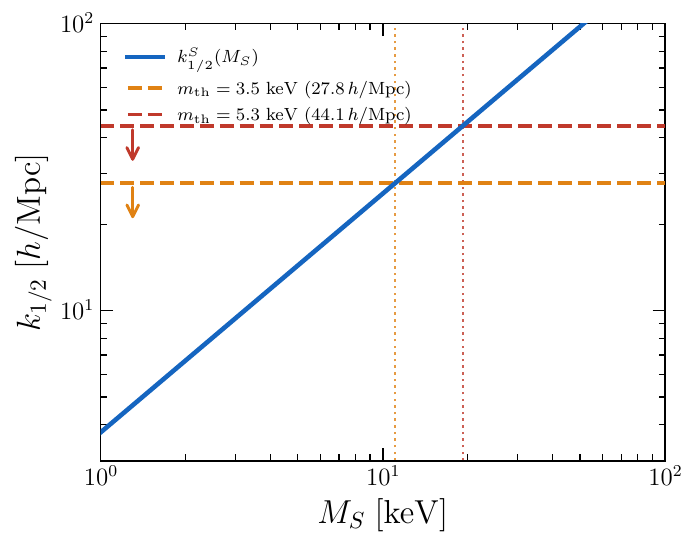}
\caption{Half-mode criterion test: $k_{1/2}(M_S)$ from
Eq.~\eqref{eq:khalf_supp} via the equivalent-mass mapping
Eq.~\eqref{eq:mth_eq}, compared to the conservative ($m_{\rm th}=3.5$~keV,
orange) and stringent ($5.3$~keV~\cite{Irsic:2017ixq}, red) reference
half-modes, both evaluated within the same fitting family.
The downward arrows indicate the excluded region,
$k_{1/2}<k_{1/2}^{\rm WDM}$.
Crossings at $M_S=11.0$ and $19.2$~keV.}
\label{fig:khalf}
\end{figure}

\section{Freeze-in production of $S$ from the same portal}
\label{sec:freezein}

The Yukawa portal $y_\alpha\,h^+\,\overline{\ell_{R\alpha}^c}\,S$,
Eq.~\eqref{eq:Ldark}, that drives DMC also produces $S$ through ordinary
freeze-in, via the decay $h^+\to\ell_\alpha^+S$ of thermal-bath charged scalars
(dominant) and the $2\to2$ scattering $\ell^+\ell^-\to S\bar S$ (subleading).
This section derives the decay yield in closed form and compares it, at the
\emph{same} benchmark parameters, with the DMC yield of the main text.

\subsection{Interactions and the two production channels}
\begin{figure}[h!]
\centering
\begin{minipage}{0.42\textwidth}
\centering
\begin{tikzpicture}
\begin{feynman}
\vertex (i1) {\(h^+\)};
\vertex[right=1.6cm of i1] (v);
\vertex[above right=1.1cm of v] (o1) {\(\ell_\alpha^{+}\)};
\vertex[below right=1.1cm of v] (o2) {\(S\)};
\diagram* {
(i1) -- [charged scalar] (v),
(v) -- [fermion] (o1),
(v) -- [fermion] (o2),
};
\end{feynman}
\end{tikzpicture}
\\[4pt]\small (A) Decay $h^+\to\ell_\alpha^{+}S$, $\;\Gamma\propto|y_\alpha|^2$.
\end{minipage}
\hfill
\begin{minipage}{0.52\textwidth}
\centering
\begin{tikzpicture}
\begin{feynman}
\vertex (i1) {\(\ell^{-}\)};
\vertex[below=1.6cm of i1] (i2) {\(\ell^{+}\)};
\vertex[right=1.6cm of i1] (a);
\vertex[right=1.6cm of i2] (b);
\vertex[right=1.1cm of a] (o1) {\(S\)};
\vertex[right=1.1cm of b] (o2) {\(\bar S\)};
\diagram* {
(i1) -- [fermion] (a) -- [fermion] (o1),
(i2) -- [anti fermion] (b) -- [anti fermion] (o2),
(a) -- [charged scalar, edge label=\(h^+\)] (b),
};
\end{feynman}
\end{tikzpicture}
\\[4pt]\small (B) $t$-channel $\ell^+\ell^-\to S\bar S$, $\;\sigma\propto|y|^4$.
\end{minipage}
\caption{Freeze-in production of $S$ from the portal Eq.~\eqref{eq:Ldark}.
(A) The dominant decay channel.
(B) The $2\to2$ scattering, with a $u$-channel partner understood.}
\label{fig:freezein_diagrams}
\end{figure}
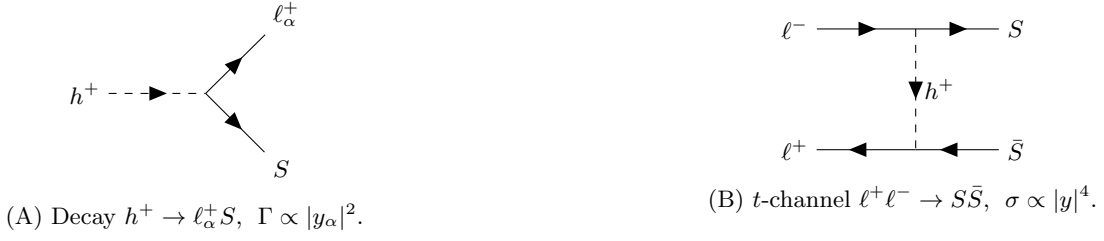

In the DMC mechanism the portal, evaluated in the domain-wall background where
$h^+$ condenses ($\langle h^+\rangle=v_+(x)\neq0$ in the core), gives a
space-dependent Dirac mass $M(x)=y_\alpha v_+(x)$ and a \emph{coherent}
wall-crossing conversion $\ell\to S$ with probability $\mathcal{P}=\tanh^2k$,
$k=y_\alpha\mathcal{S}$ (Eq.~\eqref{eq:Pgeneric}).
That is the leading channel, treated in the main text.
Here we compute instead the incoherent, homogeneous freeze-in production that
proceeds in the bulk plasma from the free $h^+$ and $\ell$ quanta, a
contribution that is always present and must be added to the DMC yield.
Channel (A) is the decay $h^+\to\ell_\alpha^+S$
(Fig.~\ref{fig:freezein_diagrams}A), a renormalizable, IR-dominated process with
most of the abundance built up at $T\sim m_{h^+}$.
Channel (B) is the annihilation $\ell^+\ell^-\to S\bar S$ via $t/u$-channel
$h^+$ exchange (Fig.~\ref{fig:freezein_diagrams}B), shown in
Sec.~\ref{sm:channelB} to be negligible while the decay is open.
Throughout, $S$ never thermalizes ($f_S\simeq0$): its only coupling is the tiny
$y_\alpha\sim10^{-9}$--$10^{-12}$.

\subsection{Freeze-in Boltzmann framework}
Let $n_S$ be the number density of $S+\bar S$ and $Y_S\equiv n_S/s$ its yield,
with $s=\tfrac{2\pi^2}{45}g_{*s}T^3$.
In a radiation-dominated universe, $H=1.66\sqrt{g_*}\,T^2/M_{\rm pl}$, and with
$x\equiv m_{h^+}/T$,
\begin{equation}
\frac{dY_S}{dx}=\frac{\mathcal{C}(T)}{s\,H\,x},
\label{eq:freezein_master}
\end{equation}
where $\mathcal{C}(T)$ is the collision term, evaluated in the
Maxwell-Boltzmann approximation with $f_S\simeq0$.

\subsection{Channel A: decay freeze-in $h^+\to\ell_\alpha^\pm S$}
From Eq.~\eqref{eq:Ldark}, the spin-summed amplitude for
$h^+(p_1)\to\ell^+(p_2)\,S(p_3)$ gives, for a massless lepton,
\begin{equation}
\Gamma(h^+\to\ell_\alpha^+S)
=\frac{|y_\alpha|^2}{16\pi}\,m_{h^+}\left(1-\frac{M_S^2}{m_{h^+}^2}\right)^{\!2}
\xrightarrow[M_S\ll m_{h^+}]{}\frac{|y_\alpha|^2}{16\pi}\,m_{h^+},
\label{eq:freezein_Gamma}
\end{equation}
and, summing flavors,
$\Gamma_{h^+}=\big(\sum_\alpha|y_\alpha|^2\big)m_{h^+}/16\pi$.
Note that the phase-space factor vanishes continuously as $M_S\to m_{h^+}$, so
the coupling required to reproduce the observed abundance through this channel
diverges at the kinematic boundary rather than jumping by a finite amount.
For a decaying bath particle with $g_{h^+}=2$ dof, the two charge states, the
collision term is
\begin{equation}
\mathcal{C}_{\rm dec}(T)=\frac{g_{h^+}\,m_{h^+}^2\,T}{2\pi^2}\,\Gamma_{h^+}\,
K_1\!\Big(\frac{m_{h^+}}{T}\Big).
\label{eq:freezein_Cdec}
\end{equation}
Inserting into Eq.~\eqref{eq:freezein_master} with $x=m_{h^+}/T$ and using
$\int_0^\infty x^3K_1(x)\,dx=3\pi/2$, with $g_{*s}\sqrt{g_*}$ slowly varying
over the production window (dominated by $T\sim m_{h^+}$),
\begin{equation}
Y_S^{\rm dec}
=\frac{135\,g_{h^+}}{1.66\times128\,\pi^4}\,
\frac{M_{\rm pl}\,\sum_\alpha|y_\alpha|^2}{g_{*s}\sqrt{g_*}\;m_{h^+}}
\simeq1.31\times10^{-2}\;\frac{M_{\rm pl}\,\sum_\alpha|y_\alpha|^2}{g_{*s}\sqrt{g_*}\;m_{h^+}}.
\label{eq:freezein_Ydec}
\end{equation}
Because Eq.~\eqref{eq:Ldark} is renormalizable, the integrand $x^3K_1(x)$ peaks
at $x\simeq2.4$ ($T\simeq0.4\,m_{h^+}$): the abundance is IR-dominated,
insensitive to the highest temperatures, and independent of the reheating
history provided $T_{\rm RH}\gtrsim m_{h^+}$.

\subsection{Channel B: annihilation freeze-in}
\label{sm:channelB}
The reaction proceeds by $t/u$-channel $h^+$ exchange with
$\sigma\propto\big(\sum_\alpha|y_\alpha|^2\big)^2/s$.
Comparing the production \emph{rates} at the common dominant temperature
$T\sim m_{h^+}$,
\begin{equation}
\frac{\mathcal{C}_{\rm ann}}{\mathcal{C}_{\rm dec}}\sim
\sum_\alpha|y_\alpha|^2\Big(\frac{T}{m_{h^+}}\Big)^4\Bigg|_{T\sim m_{h^+}}
\sim\sum_\alpha|y_\alpha|^2 ,
\end{equation}
so with $\sum_\alpha|y_\alpha|^2\sim10^{-16}$--$10^{-23}$ across the viable
range, channel B is negligible \emph{while the decay is open} and
$Y_S^{\rm FI}\simeq Y_S^{\rm dec}$.
Above $M_S=m_{h^+}$ the decay closes and channel B becomes the only surviving
freeze-in process.
Its yield is then quadratic in $\sum_\alpha|y_\alpha|^2$ and requires two bath
particles above threshold, costing $e^{-2M_S/T}$, against the single exponential
of the wall-crossing flux.
This is the origin of the break in the dotted curves of
Fig.~\ref{fig:fig_mS_sumY}.

\subsection{Comparison with the DMC yield at the same benchmarks}
The DMC yield of the main text (unsuppressed regime $k\ll1$,
$\tanh^2k\to k^2$) is, from Eq.~\eqref{eq:yield} with $\gamma_w=1$,
\begin{equation}
Y_S=\hat{\mathcal{C}}\,\frac{m_\sigma^0}{T_c^2}\,\mathcal{J}\,\sum_\alpha|y_\alpha|^2,
\qquad
\hat{\mathcal{C}}=\frac{45\,g_{\rm dof}\zeta_3}{2(1.66)\pi^4}M_{\rm pl}\simeq3.1\times10^{18}~{\rm GeV}.
\end{equation}
Both $Y_S$ and $Y_S^{\rm FI}$ are $\propto\sum_\alpha|y_\alpha|^2$, since the
portal factorizes out, and $\propto M_{\rm pl}$, in DMC through the wall area
and in freeze-in through the Hubble rate, so
\begin{equation}
\frac{Y_S}{Y_S^{\rm FI}}
=\frac{\hat{\mathcal{C}}\,(m_\sigma^0/T_c^2)\,\mathcal{J}}
{1.31\times10^{-2}\,M_{\rm pl}/(g_{*s}\sqrt{g_*}\,m_{h^+})}
\label{eq:freezein_ratio}
\end{equation}
is independent of $\sum_\alpha|y_\alpha|^2$ and of $M_{\rm pl}$, a pure function
of the wall and thermal parameters, evaluated numerically with the full running
$g_*(T)$ in Table~\ref{tab:freezein_compare}.
Because $m_{h^+}$, $m_\sigma^0$ and $T_c$ all scale $\propto v_\sigma$, and (for thin walls)
$\mathcal{S}$ is a pure function of the dimensionless quartics,
Eq.~\eqref{eq:freezein_ratio} is roughly $v_\sigma$-independent: the freeze-in
correction neither grows nor shrinks appreciably as one moves up the mass axis.
The ratio scales linearly with $\gamma_w$, so a network with $\gamma_w<1$ reduces
the entries of Table~\ref{tab:freezein_compare} proportionally.
\begin{table}[h]
\centering
\caption{DMC vs.\ freeze-in comparison for BP1--BP6, $\gamma_w=1$.
$K_0\equiv Y_S^{\rm FI}/\sum_\alpha|y_\alpha|^2$ from the full numerical
Boltzmann evolution of Eq.~\eqref{eq:freezein_master} with running $g_*(T)$;
$Y_S/Y_S^{\rm FI}$ from Eq.~\eqref{eq:freezein_ratio} at the same
$\sum_\alpha|y_\alpha|^2$, so it is independent of the coupling.}
\label{tab:freezein_compare}
\begin{ruledtabular}
\begin{tabular}{lcccccc}
BP & $v_\sigma$ [GeV] & $m_{h^+}$ [GeV] & $T_c$ [GeV] & $T_{\rm onset}$ [GeV] &
$K_0$ [GeV$^{-1}$] & $Y_S/Y_S^{\rm FI}$\\
\hline
BP1 & 300   & 523  & 350   & 263   & $3.05\times10^{11}$ & 103 \\
BP2 & 500   & 542  & 660   & 463   & $2.93\times10^{11}$ & 67  \\
BP3 & 800   & 579  & 1119  & 762   & $2.71\times10^{11}$ & 51  \\
BP4 & 2000  & 789  & 3329  & 2545  & $1.91\times10^{11}$ & 32  \\
BP5 & 5000  & 2399 & 7575  & 5691  & $6.02\times10^{10}$ & 49  \\
BP6 & 10000 & 4747 & 15150 & 11449 & $3.04\times10^{10}$ & 53  \\
\end{tabular}
\end{ruledtabular}
\end{table}

\paragraph{No double counting.}
DMC, the coherent conversion on the wall, is active for
$T_{\rm ann}<T<T_{\rm onset}$, whereas freeze-in, the incoherent bulk
production, is dominated at $T\sim m_{h^+}\gg T_{\rm onset}$ for our benchmarks.
The two populate different epochs and are additive,
$Y_S^{\rm tot}=Y_S+Y_S^{\rm FI}$.
Since $Y_S/Y_S^{\rm FI}\gg1$ throughout Table~\ref{tab:freezein_compare},
freeze-in is a $\lesssim3\%$ correction to the total abundance at the benchmarks
used in the main text, and DMC is the leading production mechanism, boosted over
freeze-in by the coherent wall-area enhancement.

\section{Charged scalar decay width and collider phenomenology}
\label{sec:llp_sm}

Using the partial width $\Gamma(h^+\to\ell_\alpha^+S)$ of
Eq.~\eqref{eq:freezein_Gamma} and its flavor sum $\Gamma_{h^+}$
(Sec.~\ref{sec:freezein}, flavor-democratic couplings, $\mathcal{B}\simeq1/3$
each), the proper decay length follows from restoring
$\hbar c\simeq1.973\times10^{-16}$~GeV$\cdot$m,
\begin{equation}
c\tau=\frac{16\pi\,\hbar c}{\big(\sum_\alpha|y_\alpha|^2\big)\,m_{h^+}}
\simeq\frac{9.9\times10^{-15}\;{\rm GeV\!\cdot\!m}}{\big(\sum_\alpha|y_\alpha|^2\big)\,m_{h^+}}.
\label{eq:llp_ctau}
\end{equation}
Table~\ref{tab:llp_bench} evaluates $c\tau$ for BP1--BP6
(Table~\ref{tab:bench}), at their own $m_{h^+}$, using the couplings
$\sum_\alpha|y_\alpha|^2$ required by Eq.~\eqref{eq:master} for two
representative masses: the Lyman-$\alpha$ floor ($M_S=3$~keV) and a
representative heavy point, $M_S=T_{\rm onset}/10$.

\begin{table}[h]
\centering
\caption{Proper decay length $c\tau$, Eq.~\eqref{eq:llp_ctau}, for BP1--BP6 at
two representative points of the viable mass range.}
\label{tab:llp_bench}
\begin{ruledtabular}
\begin{tabular}{lccc}
BP & $m_{h^+}$ [GeV] & $c\tau|_{M_S=11\,{\rm keV}}$ & $c\tau|_{M_S=T_{\rm onset}/10}$\\
\hline
BP1 & 523  & 14.7~m   & $3.6\times10^7$~m \\
BP2 & 542  & 8.8~m    & $3.8\times10^7$~m \\
BP3 & 579  & 5.9~m    & $4.1\times10^7$~m \\
BP4 & 789  & 1.8~m    & $4.4\times10^7$~m \\
BP5 & 2399 & 30~cm    & $1.6\times10^7$~m \\
BP6 & 4747 & 8.4~cm   & $8.7\times10^6$~m \\
\end{tabular}
\end{ruledtabular}
\end{table}

At the Lyman-$\alpha$ floor, $c\tau$ spans centimeters to tens of meters across
BP1--BP6.
The heavier benchmarks BP5 and BP6, with $c\tau$ of order $10$~cm, give the
pair-production topology $pp\to h^+h^-\to\ell_\alpha^+\ell_\beta^-+2S$ inside
the displaced-lepton acceptance window used by ATLAS long-lived-slepton
searches~\cite{ATLAS:2024kqk}, with $\tilde\ell\to\ell+\tilde G$ and $\tilde G$
playing the role of $S$.
For BP1--BP4 the decay length exceeds the detector scale and $h^+$ instead
traverses the tracker as a heavy stable charged
particle~\cite{Fairbairn:2006gg}.
At $M_S=T_{\rm onset}/10$ the couplings are tiny enough that $c\tau$ is
astronomical for every benchmark: $h^+$ is detector-stable and subject to HSCP
searches.
Sharing the Drell-Yan cross section of a right-handed stau, its production is
constrained by the CMS Run~2 HSCP bound
$m_{\tilde\tau_R}\gtrsim0.52$~TeV~\cite{CMS:2024nhn}, weaker in the ATLAS
$\mathrm{d}E/\mathrm{d}x$ search~\cite{ATLAS:2022pib}, and flavor-independent
since $h^+$ never decays inside the detector.
BP1--BP3 sit at or below this bound, making them directly testable at the LHC.

Finally, the same decay leaves nucleosynthesis untouched.
The proper lifetime
$\tau_{h^+}\simeq3.3\times10^{-23}\,\mathrm{s}/[\sum_\alpha|y_\alpha|^2(m_{h^+}/\mathrm{GeV})]$
is longest at the heavy end, yet even there
($\sum_\alpha|y_\alpha|^2\sim10^{-26}$, $m_{h^+}=500$~GeV) it stays well before
deuterium formation and far before the $^6$Li catalysis epoch
($t\gtrsim10^3$~s)~\cite{Pospelov:2006sc,Cyburt:2015mya}.

\end{document}